\documentclass[preprint2]{aastex}

\begin{document}

\title{Distributions of Galaxy Spectral Types in the Sloan Digital Sky Survey}

\author{C.~W.~Yip\altaffilmark{1},
A.~J.~Connolly\altaffilmark{1}, 
A.~Szalay\altaffilmark{2}, 
T.~Budavari\altaffilmark{2}, 
M.~SubbaRao\altaffilmark{3}, 
J.~Frieman\altaffilmark{3}, 
R.~Nichol\altaffilmark{4},
A.~Hopkins\altaffilmark{1},
D.~York\altaffilmark{3,5},
S.~Okamura\altaffilmark{6},
J.~Brinkmann\altaffilmark{7},
I.~Csabai\altaffilmark{2,8}, 
A.~R.~Thakar\altaffilmark{2},
M.~Fukugita\altaffilmark{9},
Z.~Ivezi\'{c}\altaffilmark{10}}

\altaffiltext{1}{Department of Physics and Astronomy, University of Pittsburgh, 
Pittsburgh, PA 15260, USA} 

\altaffiltext{2}{Department of Physics and Astronomy, Johns Hopkins University,
3701 San Martin's Drive, Baltimore, MD21218, USA}

\altaffiltext{3}{Department of Astronomy
and Astrophysics, The University of Chicago, 5640 S. Ellis Ave, Chicago,
IL, 60637 USA}

\altaffiltext{4}{Department of Physics, Carnegie Mellon University, 
5000 Forbes Ave., Pittsburgh, PA 15213, USA}

\altaffiltext{5}{Enrico Fermi Institute, The University of Chicago, 
5640 South Ellis Avenue, Chicago, IL 60637, USA}

\altaffiltext{6}{Department of Astronomy and Research Center for the Early Universe,
School of Science, University of Tokyo,
Tokyo 113-0033, Japan}

\altaffiltext{7}{Apache Point Observatory, 2001 Apache Point Road, P.O. Box 59,
Sunspot, NM88349-0059, USA}

\altaffiltext{8}{Department of Physics, E\"otv\"os University, Budapest, Pf. 32, Hungary,
H-1518}

\altaffiltext{9}{Institute for Cosmic Ray Research,
University of Tokyo, Japan}

\altaffiltext{10}{2 Peyton Hall, Princeton University, Princeton, NJ 08544, USA}

\begin{abstract}
We perform an objective classification of 170,000 galaxy spectra in the
Sloan Digital Sky Survey (SDSS) using the Karhunen-Lo\`eve (KL)
transform. With about one-sixth of the total set of galaxy spectra which
will be obtained by the survey, we are able to carry out the most
extensive analysis of its kind to date. The formalism proposed by
Connolly and Szalay~(1999a) is adopted to correct for gappy regions in
the spectra, and to derive eigenspectra and eigencoefficients. From this
analysis, we show that this gap-correction formalism leads to a
converging set of eigenspectra and KL-repaired spectra.  Furthermore, KL
eigenspectra of galaxies are found to be convergent not only as a
function of iteration, but also as a function of the number of randomly
selected galaxy spectra used in the analysis. From these data a set of
ten eigenspectra of galaxy spectra are constructed, with rest-wavelength
coverage $3450-8350$~\AA. The eigencoefficients describing these galaxies
naturally place the spectra into several classes defined by the plane
formed by the first three eigencoefficients of each spectrum. Spectral
types, corresponding to different Hubble-types and galaxies with extreme
emission lines, are identified for the 170,000 spectra and are shown to
be complementary to existing spectral classifications.  From a
non-parametric classification technique, we find that the population of
galaxies can be divided into three classes which correspond to early
late- through to intermediate late-types galaxies. This finding is
believed to be related to the color separation of SDSS galaxies
discussed in earlier works. Bias in the spectral classifications due to
the aperture spectroscopy in the SDSS is small and within the
signal-to-noise limit for majority of galaxies except for the reddest nearby
galaxies and large galaxies ($>30$ kpc) with prominent emissions. The
mean spectra and eigenspectra derived from this work can be downloaded
from http://www.sdss.org.
\end{abstract}
 

\section{Introduction}

The most successful scheme used to date to classify galaxies is the
morphological classification of Hubble~\cite{Hubble26}. The utility of
this simple classification scheme (a compression of the available
morphological types to approximately seven classes) has become apparent
through its application to numerous extragalactic studies. Spectral
classifications have a number of natural advantages over the
morphological classifications of Hubble in that they are more easily
related to the physical processes that are ongoing within a galaxy
(e.g., star formation) and that they do not require us to obtain high
resolution imaging of a large number of galaxies. As such, they are well
suited to studying the cores of galaxies in the distant universe.

As was found for the classification of the spectra of stars, classifying
the spectra of more complicated systems such as galaxies or quasars
(QSOs) can provide a better understanding of the physical processes that
determine the formation and evolution of these sources.  Moreover, if
there exist mechanisms by which galaxies can be classified using a
handful of representative parameters, this classification can be thought
of as a compression of the information contained within the
spectra. From such an approach, one might be able to derive simple
mechanisms for exploring the physics of the spectral properties of
galaxies using large data sets. Recent massive spectroscopic surveys,
e.g., the Anglo-Australian Observatory 2-degree-Field (2dF) Galaxy
Survey (Colless et~al.\ 2001) and Sloan Digital Sky Survey (SDSS; York et
al.\ 2000) provide us with the opportunity to address the classification
of galaxy spectra using hundreds of thousands of galaxy spectral energy
distributions (SEDs). One technique that has gained popularity for
studying the distribution of SEDs is the Karhunen-Lo\`eve (KL)
transform. The power of this approach is that it enables a large amount
of data to be decomposed and compressed into independent components in
an objective way. Applications of this technique can be found in the
classifications of galaxies (Connolly et~al.~1995; Folkes et~al.~1996; 
Sodre \& Cuevas~1997; Bromley~et~al.~1998; Galaz \& de Lapparent~1998; 
Ronen~et~al.~1998; Folkes~et~al.~1999), 
QSOs (Francis et~al.~1992; Boroson \& Green~1992; 
Yip et~al.~2004) and stars \cite{Singh98, Bailer-Jones98}.

This paper is organized as follows. In Section~\ref{section:data}, we
describe the spectral data used in our analysis. In
Section~\ref{section:KLintro}, we discuss the details of the
Karhunen-Lo\`eve transform and the gap-correction formalism.  In
Section~\ref{section:conv}, we pose the problems to be addressed with
this paper, and show the results of a convergence analysis on the KL
gap-correction formalism. In Section~\ref{section:KL}, we derive the
eigenspectra and eigencoefficients for the full SDSS data set, and
implement a classification scheme. In Section~\ref{section:repeat}, we
discuss the reliability of this classification.  A simple model is used
to describe the population of galaxies in Section~\ref{section:model}.
In Section~\ref{section:app}, we discuss the applications of the KL
eigenspectra obtained in this work. In
Section~\ref{section:aperturebias}, we discuss the aperture bias effect
on the current classification scheme. Finally, in
Section~\ref{section:conclusion} we conclude our results and discuss
some possible future directions based on this work.

\section{Data} \label{section:data}
                         
As part of the Sloan Digital Sky Survey (York et~al.\ 2000) spectra are
taken with fibers of 3 arcsec diameter (corresponding to 0.18mm at the
focal plane for the 2.5m, f/5 telescope). All sources are selected from
an initial imaging survey using the SDSS camera described in Gunn et
al.\ (1998) with the filter response curves as described in Fukugita et
al.\ (1996), and using the imaging processing pipeline of Lupton et~al.\
(2000). The astrometric calibration is described in Pier et~al.\
(2002). The photometric system and monitoring are described in detail in
Smith et~al.\ (2002) and Hogg et~al.\ (2001) respectively. To date,
there are three complete samples of SDSS spectra: the Main Galaxy sample
(Strauss et~al.\ 2002), the Luminous Red Galaxy sample (LRGs; Eisenstein
et~al.\ 2002), and the QSO sample (Richards et~al.\ 2002). From these
data we select the Main Galaxy sample for our analysis and use only
those galaxies defined as being of survey quality: a signal-to-noise
lower-limit of approximately 16.  The galaxies in this sample have
r-band Petrosian magnitudes $r_p < 17.77$ and Petrosian half-light
surface brightnesses $\mu_{50} < 24.5$ mag arcsec$^{-2}$, defined to be
the mean surface brightness within a circular aperture containing half
of the Petrosian flux (called the Petrosian half-light radius). The
spectral reductions used are the standard SDSS 2D analysis pipeline
(idlspec2D v4.9.8, as of 18th of April, 2002) and the 1D SpecBS pipeline
(Schlegel et~al.\ 2003). The resultant spectra are flux- and
wavelength-calibrated, and sky-subtracted. From these data approximately
two hundred galaxies are removed, as they have zero flux in all
pixels. This results in a final sample of 176,956 galaxy spectra. The
median redshift is about 0.1, and we find that about 6.5\% of the sample
have redshifts $cz<10,000$~kms$^{-1}$, so that their Petrosian
half-light radii can be substantially larger than the 3 arcsec aperture
of the fiber (Strauss et~al.\ 2002).  1,854 spectra of the final sample
are found to be duplicated observations; identified as being within a
search radius of 2 arcsec and a redshift tolerance of 0.01. All spectra
are shifted to a common rest frame, and rebinned to a vacuum wavelength
coverage of $3450 - 8350$~\AA. The binning of the spectra is logarithmic,
with a velocity dispersion of 69~kms$^{-1}$.  This procedure emphasizes
the blue end of the optical spectrum, enabling our analysis to focus on
the Ca~H and Ca~K lines, and the Balmer break. The resultant spectra cover
rest-wavelength range $3450 - 8350$~\AA\ over 3839 pixels.

\section{KL and Gap-Correction Formalism} \label{section:KLintro}

The Karhunen-Lo\`eve transform (or Principal Component Analysis, PCA) is
a powerful technique used in classification and dimensional reduction of
data. In astronomy, its applications in studies of multivariate
distributions have been discussed in detail (Efstathiou \& Fall~1984; Murtagh \& Heck~1987).  
In this paper, we limit ourselves to its applications to
spectral energy distributions. The basic idea is to derive a lower
dimensional set of {\it eigenspectra} \cite{Con95} from a very large set
of input SEDs.  Each SED can be thought of as an axis in a
multidimensional hyperspace, $f_{\lambda_{k} i}$, where $\lambda_{k}$
denotes the $k$-th wavelength in the $i$-th galaxy spectrum.

For the moment, we assume that there are no gaps in each spectrum; we
will discuss the ways we deal with gappy regions later.  From the set of
spectra we construct the correlation matrix
\begin{equation}
  C_{{\lambda_{k}}{\lambda_{l}}} = \hat{f}_{\lambda_{k} i} \hat{f}_{i \lambda_{l}} \ ,
\end{equation}

\noindent
where the summation is from $i=1$ to the total number of spectra, and
$\hat{f}_{\lambda_{k} i}$ is the normalized $i$-th spectrum, defined for
a given $i$ as,
\begin{equation}
  \hat{f}_{\lambda_{k}} = {{f_{\lambda_{k}}}\over{\sqrt{f_{\lambda_{k}} \cdot f_{\lambda_{k}}}}}\ .
\end{equation}

The eigenspectra are obtained by finding a matrix, $U$, such that
\begin{equation}
 U^{T} C U = \Lambda \ ,
\end{equation}

\noindent
where $\Lambda$ is the diagonal matrix containing the eigenvalues of
the correlation matrix. $U$ is thus a matrix whose $i$-th column
consists of the $i$-th eigenspectrum ${e}_{i {\lambda_{k}}}$.  We
solve this eigenvalue problem by using Singular Value Decomposition.

The observed spectra are projected on to the eigenspectra to obtain the
eigencoefficients. In these projections, every pixel in each spectrum is
weighted by the error associated with that particular pixel,
$\sigma_{\lambda}$, such that the weights are given by
$w_{\lambda}=1/{\sigma_{\lambda}}^2$. The observed spectra can be
decomposed, with no error, as follows
\begin{equation}
 {f}_{\lambda_{k}} = \sum_{i=1}^{M} a_{i} e_{i {\lambda_{k}}}\ ,
\end{equation}

\noindent
where $M$ is the total number of eigenspectra. It is straightforward
to see that $M$ equals the total number of wavelength bins in the 
spectrum.

Previously, we assumed that the spectra are without any gaps.  In
reality, however, there are several reasons for gaps to exist: for
example, different rest-wavelength coverages, the removal of sky lines,
bad pixels on the CCD chips etc.\ leave gaps at different rest frame
wavelengths for each spectrum. All can contribute to incomplete
spectra. The principle behind the gap-correction process is to
reconstruct the gappy spectrum using its principal components. The first
application of the method to analyze galaxy spectra is due to Connolly
and Szalay~(1999a), which expands on a formalism developed by Everson and
Sirovich for dealing with two-dimensional images
\cite{Everson94}. Initially, we fix the gap regions by some means, for
example, linear-interpolation.  A set of eigenspectra are then
constructed from the gap-repaired galaxy spectra. Afterward, the gaps in
the original spectra are corrected with the linear combination of the KL
eigenspectra. The whole process is iterated until the eigenspectra
converge, which we define in the next section. According to Everson and
Sirovich's work on artificially masked two-dimensional images, they
claimed that the iteration process gives convergent images.  However,
the question about whether the principal components resulting from the
correction procedure on realistic gappy galaxy spectra would converge is
unknown and is to be addressed with this work.

\section{A Convergence Analysis of KL} \label{section:conv}

There are some questions to be solved in our analysis before the KL
eigenspectra and hence the classification itself become robust and
meaningful. These questions are: do the resultant eigenspectra converge
and, if so, how many iteration steps are required, what is the
dependency of the quality of the KL-repaired spectra on how the gaps are
initially corrected, how much information is contained in the
eigenspectra and most importantly, how many galaxy spectra are needed in
order to derive a convergent set of eigenspectra?

Several authors have tried to assess the performance of a KL analysis in
a number of different quantitative ways. An example of this is the
$\chi^2$ assessment \cite{Francis92} in which the authors calculated the
difference between the observed spectrum and the spectrum reconstructed
with the principal components in order to determine the number of
components needed for reconstructing a quasar spectrum.  With the
implementation of gap-corrections in our analysis, this comparison of
only {\it one} spectrum to another may not suffice. We are more interested
in how well the {\it set} of eigenspectra describes the distribution
of spectra rather than a one-to-one comparison.  For example, how does
the set of eigenspectra differ as the gap-correction procedure
progresses?  Given two subspaces, each formed by a set of eigenspectra
obtained with different conditions (e.g., at different points in the
iterative gap correction or computed with different numbers of observed
spectra), we require a method that will quantitatively compare one set
of eigenspectra with another. In other words, instead of just comparing
two spectra, a mechanism is desired to compare two subspaces, which are
spanned by a finite number of spectra respectively. Mathematically it
can be stated, as in \cite{Everson94}, that two spaces, $E$ and $F$, are
in common if
\begin{eqnarray}
 Tr(\mathbf{EFE}) & = & D \ ,
\end{eqnarray} 

\noindent 
where $\mathbf{E}$ and $\mathbf{F}$ are the sum of projection
operators of space $E$ and $F$ respectively, and $D$ is equal to the
dimensionality of each space. We assume that these eigenbases have the
same dimensions for a meaningful comparison.  The sum of the
projection operators, $\mathbf{E}$, of a space is given by the sum of
the outer products
\begin{eqnarray}
 \mathbf{E} = \sum_{\epsilon} |\epsilon> <\epsilon| \ , 
\end{eqnarray} 

\noindent 
where $|\epsilon>$ are the basis vectors which span the space $E$ (e.g.,
Merzbacher 1970). A basis vector is an eigenspectrum if $E$ is
considered to be a set of eigenspectra.  The two spaces are disjoint if
the trace quantity is zero and are identical if the quantity is equal to
the dimension of the subspace.  This provides a quantitative way of
measuring the commonality of two subspaces (i.e., how similar the two subspaces are).

\begin{figure}[h]
\includegraphics[angle=270,width=3in]{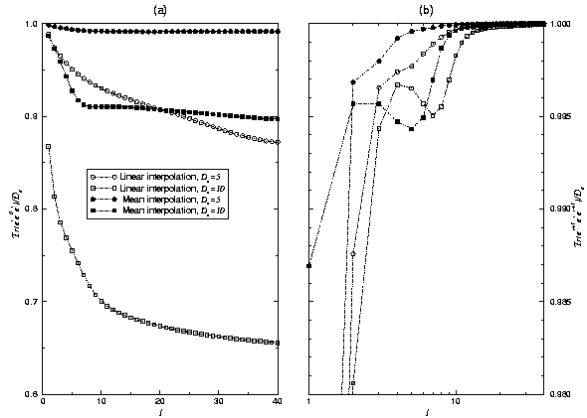}
\figcaption[eigentrace_10_all.eps]{Convergence of eigenspectra (a)
	$Tr(e^{i}e^{0}e^{i})$ and (b) $Tr(e^{i+1}e^{i}e^{i+1})$ as a
	function of iteration step in the KL gap-correction
	formalism.  Curves with open symbols are
	linearly-interpolated across the gaps while filled symbols
	represent mean-interpolated data.  The circles and squares denote
	$D_e=5$ and $10$ respectively, where $D_e$ is the dimension of
	the subspace formed by the eigenspectra.  \label{fig:eigspecTraceDe}}
\end{figure}

In investigating the convergence behavior of eigenspectra as a function
of the number of iterations, we define one of the spaces to be that
formed by a finite number of eigenspectra obtained after initially
interpolating over the gap regions, but {\it without} gap
correction. The other spaces are defined to be those formed by the same
finite number of eigenspectra but at different iterations in the gap
correction. The sum of projection operators in the first case is named
$e^{0}$ and in the latter, $e^{i}$. The subspaces are named $E^{0}$ and
$E^{i}$ respectively, where $i$ denotes the $i$-th iteration. The
dimension of the space is $D_e$, which is the number of eigenspectra
forming the subspace.

The trace quantity $Tr(e^{i}e^{0}e^{i})$ as a function of iteration is
plotted in Figure~\ref{fig:eigspecTraceDe}a, in which the KL transform
is applied to $N=4003$ randomly-chosen SDSS galaxy spectra, where $D_e$
is set to be $5$ and $10$, meaning that the subspaces are spanned by the
the first five and the first ten eigenspectra respectively. Repairing of
the galaxy spectra in the iteration procedure is performed with $m=10$
eigenspectra.  It should be noted that $m$ is independent from $D_e$ and
that $D_e$ is always smaller or equal to $m$. The traces are normalized
by the corresponding $D_e$ in each curve to simplify comparison.
Initially, let us concentrate on the curves with open symbols in
Figure~\ref{fig:eigspecTraceDe}.  These curves denote that the initial
gappy regions are approximated by linear-interpolation. In this
linear-interpolation method the flux of a pixel, $f^{g}_{\lambda_{k}}$,
in the gap is simply approximated by the average of its neighbors, so that
\begin{equation}
f^{g}_{\lambda_{k}} = (f_{\lambda_{k-1}}+f_{\lambda_{k+1}})/2 \ .
\end{equation}

The trace quantity decreases gradually as the iteration step increases,
indicating that the space $E^{i}$ is less and less in common with
$E^{0}$. This implies that the KL gap-correction and eigenspectra
construction are changing the spectrum of a galaxy within the gappy
regions. As such the eigenspectra from the KL-repaired spectra differ
progressively more from those formed from the original spectra. The
above is true for both $D_e = 5$ and $10$. The above is generally valid
for $D_e$ from $1-10$. As the iteration increases, the slope of the
curve decreases, which implies that a we have converging set of
eigenspectra.

The choice of linear-interpolation in the initial correction for the
gappy regions is arbitrary. In fact, if the gap formalism is robust, the
quality of the KL-repaired spectrum and the eigenspectra should be
independent of the way the observed spectra are initially repaired. We
test an alternative method of correcting for gaps, where the flux at
each wavelength bin in the gap region is approximated by the mean of all
other spectra within that region, i.e., the flux of a pixel in a gap is
approximated by,
\begin{equation}
\hat{f}^{g}_{\lambda_{k}} = \left< \sum_{\mathrm all \, spec} 
\hat{f}_{\lambda_{k}} \right>_{\mathrm all \, spec} \ ,
\end{equation}

\noindent
where $\hat{f}_{\lambda_{k}}$ is the normalized flux at
${\lambda_{k}}$. We call this method mean-interpolation.  With this
alternate method the trace quantity also converges as we can see from
Figure~\ref{fig:eigspecTraceDe}a, but at a higher value than those in
the case of linear-interpolation. This behavior shows that, in the case
of mean-interpolation, the eigenspectra constructed after the gap
correction deviate {\it less} from the initial interpolated spectra than
in the case of linear-interpolation. The rate of convergence is faster
when using the mean-interpolation method. This suggests that the
mean-interpolation provides a better initial estimate of the true
spectra within the gap regions. We do not, therefore, require as many
iterations as in the case of linear-interpolation. This is important as
each step in repairing the spectra and constructing the eigenspectra is
computationally expensive when large amounts of data are under
consideration.

Figure~\ref{fig:eigspecTraceDe}b shows the convergence behavior of the
sets of eigenspectra given in Figure~\ref{fig:eigspecTraceDe}a except
that the trace quantity now compares the subspace from one iteration to
the subspace from the next iteration, i.e., $Tr(e^{i+1}e^{i}e^{i+1})$. As
expected, the convergence with the number of iteration steps can be seen
in both methods, but with this more sensitive measurement the
convergence is now no longer found to be monotonic. This implies that we
may need more iterations, than it first appeared from our previous
example in order to obtain a convergent set of eigenspectra.  Again, the
mean-interpolation method is shown to converge to a consistent set of
eigenspectra faster than for linear-interpolation. Consequently, in the
following all gaps in the spectra will be fixed using the
mean-interpolation method, unless otherwise specified.

\begin{figure}[h]
\includegraphics[angle=270,width=3in]{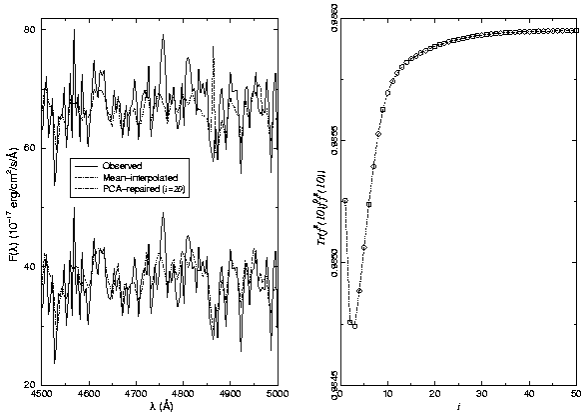}
\figcaption[Spec_1800_MI.eps]{(a) The mean-interpolated and
	 KL-repaired (20 iterations) of an artificially masked spectra, overlaid on the 
	original unmasked spectrum.  (b) KL-repaired spectra converge as a
	 function of iteration.  \label{fig:MIRepairedSpec}}
\end{figure}

An example of the actual performance in the mean-interpolated and
repaired spectra is shown in Figure~\ref{fig:MIRepairedSpec}.  The data
set is the same $4003$ galaxies as before, except that one randomly
chosen spectrum is artificially masked in the region [4500, 5000]\AA.
The upper panel of Figure~\ref{fig:MIRepairedSpec}a shows that the
mean-interpolated region is already close to that of the original
spectrum before masking. The lower panel shows the KL-repaired spectrum
at $i=20$ overlaid on the original spectrum, using all $10$
eigenspectra.  The spectra are offset by an arbitrary amount for
illustration. There is a substantial improvement in retrieving the
original spectrum as the iterations proceed. To compare the KL-repaired
and unmasked spectrum quantitatively, we apply a similar convergence
measure as described previously.  The convergence measure in this case
is defined to be
\begin{equation}
 Tr(f^{R}(m)f^{0}f^{R}(m)) \ ,
\end{equation} 

\noindent
where $f^{R}(m)$ is the projector of the KL-repaired spectrum with $m$
eigenspectra in the gaps, and $f^{0}$ is that of the unmasked
spectrum. In Figure~\ref{fig:MIRepairedSpec}b, the trace quantity versus
the number of iterations is plotted for the case corresponding to
Figure~\ref{fig:MIRepairedSpec}a.  $Tr(f^{R}(m)f^{0}f^{R}(m))=1$ means
that the repaired spectrum is identical with the original unmasked
one. We see that after the initial few iterations, the two become more
similar. After 30 iterations, the KL-repaired spectrum converges to that
of the original spectrum with a high degree of accuracy, the difference
in the trace quantities of $8\times 10^{-4}$~\%.

\begin{figure}[h]
\includegraphics[width=3in]{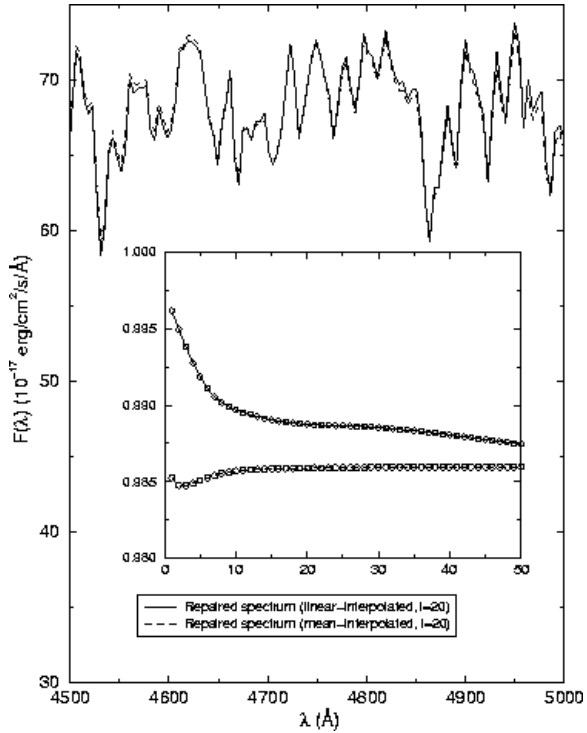}
\figcaption[repairedspec_MILI_iter20.eps]{The KL-repaired spectra,
	linear-interpolated (solid line) and mean-interpolated (broken
	line), in the gap regions of the unmasked spectrum as in
	Figure~\ref{fig:MIRepairedSpec} (after 20 iterations).  The
	insert is $Tr(f^{R}(m)f^{0}f^{R}(m))$ as a function of
	iteration step for both cases.  \label{fig:MILIrepairedspec}}
\end{figure}

Combining the results of the convergence measures in both KL-repaired
spectra and the previously discussed eigenspectra, it can be concluded
that the convergence of the eigenspectra implies the convergence of the
repaired spectrum, and vise versa. Furthermore, the quality of a
repaired spectrum should not depend on the initial gap-interpolation
technique.  Figure~\ref{fig:MILIrepairedspec} shows the KL-repaired
spectra, using linear- and mean-interpolations for the initial gap
approximation, $m=10$ and at $i=20$. The two are shown to be very
similar to each other. The insert shows the corresponding
$Tr(f^{R}(m)f^{0}f^{R}(m))$ as a function of iteration. The convergence
behavior seems different in both cases, nevertheless they are
approaching each other with a difference in the actual value about 0.3\%
which is small as can be seen in the plots of the spectra in the main
graph. This is a desired result, because if the whole formalism is
robust, the repaired spectrum should not be different due to different
procedures used in the initial gap-fixing.

\subsection{The Effect of Sample Size} \label{subsection:size}

\begin{figure}[h]
\includegraphics[angle=270,width=3in]{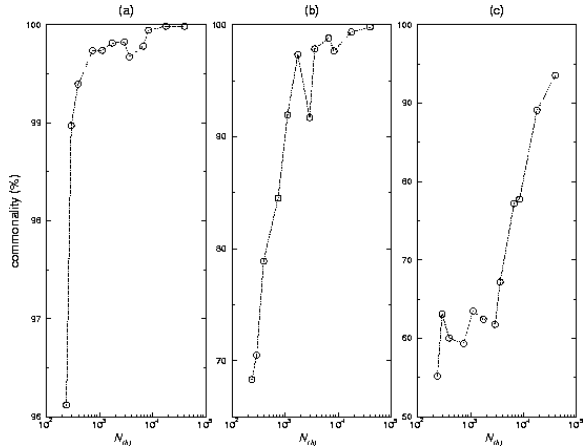}
\figcaption[eigentrace_2310_newN.eps]{The commonality measurement of
	the subspaces formed by the set of eigenspectra derived using
	different numbers of observed galaxy spectra where the
	subspaces are spanned by the first (a) 2 (b) 3 and (c) 10
	eigenspectra respectively. The results show that the
	eigenspectra set converge as a function of learning-set size
	in the KL gap-correction formalism.
	\label{fig:eigentrace_2310_newN}}
\end{figure}

Another important aspect of the KL-eigenspectra construction is the
number of observed spectra necessary as the input.  In principal, as
we increase the number of galaxy spectra, a more
representative and general set of eigenspectra should result. The question remains, however, 
how much more generality would be gained by including more observed spectra in the
analysis?  Fundamentally, does there exist a minimum number of input
galaxy spectra such that the eigenspectra set start to converge?  This
is important because we can thus use a minimum number of
randomly-chosen observed spectra in the survey to derive a set of
eigenspectra which nevertheless contain all the necessary information
within the full data set. Figure~\ref{fig:eigentrace_2310_newN} shows
an attempt to answer this question. In these figures, the commonality
percentages of two subspaces spanned by (a) $2$ (b) $3$ and (c) $10$
eigenspectra are plotted versus the number of galaxy spectra N$_{(h)}$
used in the sample. The commonality is similar to that previously
discussed for the trace quantities except here we compare the set of
eigenspectra derived from N$_{(h)}$ input galaxy spectra with that
from a smaller number of spectra N$_{(h-1)}$. This is defined as
follows
\begin{equation}
{\mathrm commonality(\%)} \equiv {{Tr(e(N_{(h-1)}) e(N_{(h)})e(N_{(h-1)})) } 
\over{D_{e}}} \times 100\% \ , 
\end{equation} 
 
\noindent
where $e(N_{(h)})$ is the sum of projectors of the subspace spanned by
$D_{e}$ eigenspectra, derived from $N_{(h)}$ galaxy spectra, using
$m=D_{e}$ eigenspectra for gap-repairing.  The number of iterations
for the gap-correction is 20.  The smallest number of spectra we
consider is $139$ (=N$_{(0)}$) and the largest number $40044$.  The
galaxies in each case are randomly selected from the full SDSS sample.

For all cases, convergent trends are present as more spectra are
included. For the case of two eigenspectra
(Figure~\ref{fig:eigentrace_2310_newN}a)  we can see that only
about $500$ galaxy spectra are needed in order to construct the first
two eigenspectra to an 0.5\% accuracy when compared to the final converged
set. The inclusion of the third mode
(Figure~\ref{fig:eigentrace_2310_newN}b) requires more spectra to
obtain a similar accuracy though the convergence behavior is very
similar to (a) (with slightly larger fluctuations). Nevertheless,
only about $1000$ galaxy spectra are needed for 90\% commonality.
These results are consistent with the fact that most types of galaxy
spectra can be described with 2 to 3 colors \cite{Con95} and
therefore a random sampling of a few thousand galaxies can be expected to cover 
the full color distribution for these galaxies.

In Figure (c), it is interesting that the convergent behavior is
different from that in (a) and (b). With a sample size smaller than
about $3000$ to $4000$, the improvement in the set of 10 eigenspectra is
small.  However, once the number of spectra used exceeds that threshold,
the convergent rate dramatically increases. This finding suggests that
there exists a minimum number of galaxy spectra that we need to include
in our KL analysis in order to fully sample the true distribution of
galaxy spectra.  Combining this with the fact that the higher-order
modes in the eigenspectra tend to correspond to spectral features in
galaxies with prominent emission lines (this will be discussed in detail
in Section~\ref{section:KL}) and the fact that those galaxies only
comprise about 0.1\% of the whole sample, the behavior in (c) can be
understood as the effect of including galaxies with relatively extreme
spectral properties. When we randomly pick about $1000$ galaxies from
the sample there are a few emission line galaxies. As more spectra are
included we eventually reach a threshold where we begin to sample the
extreme emission line galaxies. Once we have included a small number of
these emission line galaxies (with a sample size of $3000-4000$ galaxies
we would expect three to four emission line galaxies) the information
they contain is now incorporated within the KL eigenbases. The dramatic
increase in the convergence rate come from the fact that, while rare,
these extreme emission line galaxies can still be described by a handful
of spectral components (i.e., once we have a small number of them in the
sample we can map out their full distribution).

To conclude, there exist a minimum number of galaxy spectra we need to
observe in order to derive a convergent set of eigenspectra. 
Approximately $10^{4}$ spectra are sufficient for a 90\% convergence 
level with ten eigen-components (Figure~\ref{fig:eigentrace_2310_newN}c).
This is sufficient to characterize the spectral types of 99.9\% of 
galaxies within the local universe. These results are, however, purely 
empirical, based on randomly selecting spectra from the current 
data set.  Thus, there is no concrete evidence to support the present result that 
$10^{4}$ galaxy spectra are all we need in deriving the most complete set of 
eigenspectra. There may exist populations of galaxies that comprise 
much less that 0.1\% of the full galaxy sample that our current analysis 
is not sensitive to. In general, for a larger data set (e.g., at the 
completion of the survey), new galaxy types, if any, may call for more 
spectra to be included when constructing the eigenspectra. 

\section{KL Eigenspectra and ($\phi_{KL}, 
\theta_{KL}$)-Classification} \label{section:KL}

\begin{figure}[h]
\includegraphics[angle=0,width=3in]{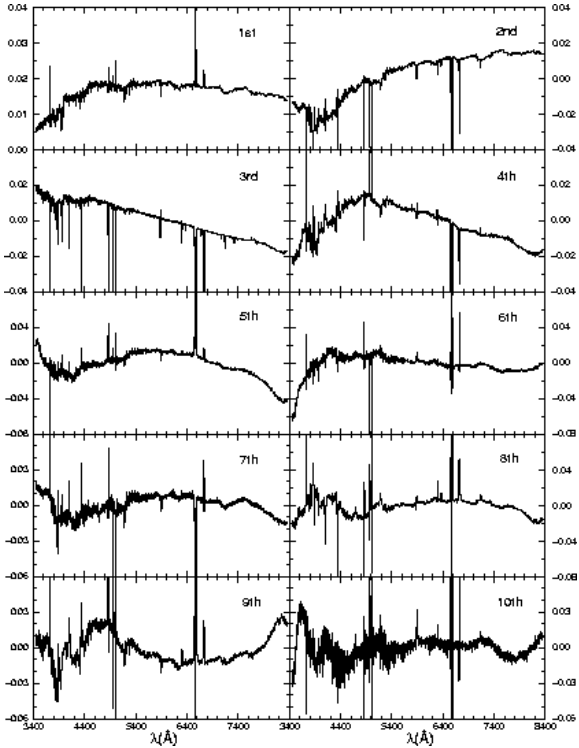}
\figcaption[eigSpec_iter20_ALL.eps]{The first 10 KL eigenspectra of
        $\approx$170,000 galaxy spectra in the SDSS. Gap correction is implemented for
        $20$ iterations. \label{fig:10eigSpec}}
\end{figure}

The first $10$ KL eigenspectra of the 170,000 SDSS galaxies are shown in
Figure~\ref{fig:10eigSpec}, derived from $20$ iterations and using 10
eigenspectra for gap repairing. The eigenspectra are publicly available
(from the website http://www.sdss.org).  The first eigenspectrum is the mean of all
galaxy spectra in our sample.  The continuum is similar to a Sb-type\setcounter{footnote}{0}
\footnote{In our work, the red- and blue-types are determined from the
spectral information in the galaxies. Thus, the conventional
morphological-type nomenclatures ``early'', ``late'', and E, S0, types
etc. used in this paper are referring to {\it spectral features} which
usually would have be seen in the corresponding morphological types, as
suggested in Kennicutt's Atlas and other studies.}.  As we would expect
from the mean of all spectra, nebular lines and other emission lines, as
well as absorption lines such as Ca~H and Ca~K, are present within this
spectrum. The second eigenspectrum has one zero crossing, positive
toward longer wavelengths, at around 5200\AA, which marks the
wavelength at which the modulation in the continuum level relative to
the 1st eigenspectrum occurs.  In the third component, there is a zero
crossing in the continuum, negative toward longer wavelengths, at around 6000\AA. The
higher the order of the eigenspectrum, the larger the number of zero
crossings which in turn adds high-frequency features to the final
spectrum as these higher order modes are added or subtracted. In the
higher-order modes, the eigenspectra are dominated by emission and
absorption lines because each of these eigenspectra comprises emission
and absorption lines plus a small fluctuation of the continuum level
around zero. We illustrate this point later in the paper.

Statistically, the amount of information contained in each
eigenspectrum is given by the eigenvalue of the correlation matrix of
that particular mode. Table~\ref{tab:weight} lists the weights
of the first $m$-modes of eigenspectra, where the weights are normalized to unity. 
We find that the first three eigenspectra
contain more than $98\%$ of the total variance of the data set.

It is known that there is a one-parameter description of galaxy
spectra which correlates with the spectral type of a
galaxy \cite{Con95}. This parameter, $\phi_{KL}$, is the mixing angle
of the first and second eigencoefficients. Explicitly,
\begin{equation}
  \phi_{KL} = \tan^{-1}(a_2/a_1) \,
\end{equation}

\noindent 
where $a_1$ and $a_2$ are the eigencoefficients of the first and
second modes of a galaxy respectively. Furthermore, the inclusion of
the third component discriminates the post-starburst activity \cite{Con95, Castander01}. 
To follow this classification scheme, we define here
\begin{equation}
  \theta_{KL} = \cos^{-1}(a_3) \,
\end{equation}

\noindent 
where $a_3$ is the third eigencoefficient. Here we adopt the
normalization 
\begin{equation}
  \sum_{k=1}^{10} {a_{k}^{2}} = 1 \ .
\end{equation}

\begin{figure}[h]
\includegraphics[angle=0,width=3in]{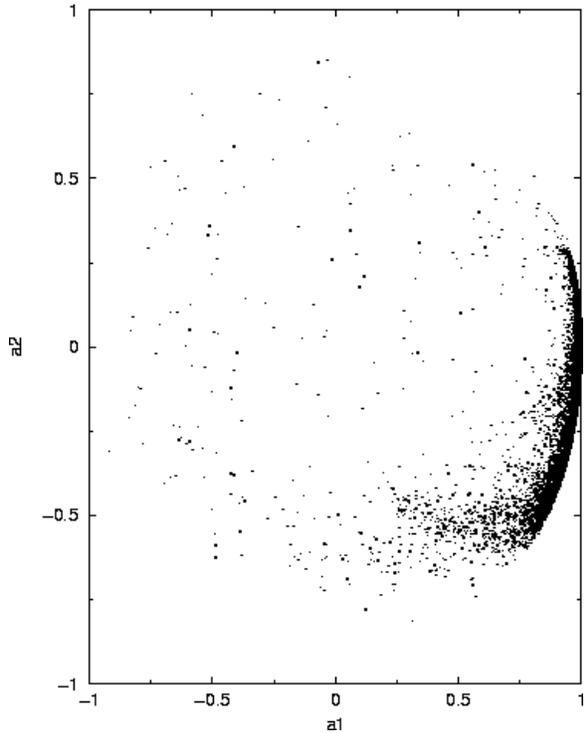}
\figcaption[a2vsa1.eps]{Eigencoefficients $a_2$ versus $a_1$ of our
sample ($\approx$170,000 galaxy spectra).  More than 90\% of the whole galaxy
population are located on this main locus. The trend is similar to that
in previous works \cite{Con95}, with red galaxies having larger,
positive $a_2$ values and blue galaxies having smaller, or negative
values.  \label{fig:a2vsa1}}
\end{figure}

\begin{figure}[h]
\includegraphics[angle=0,width=3in]{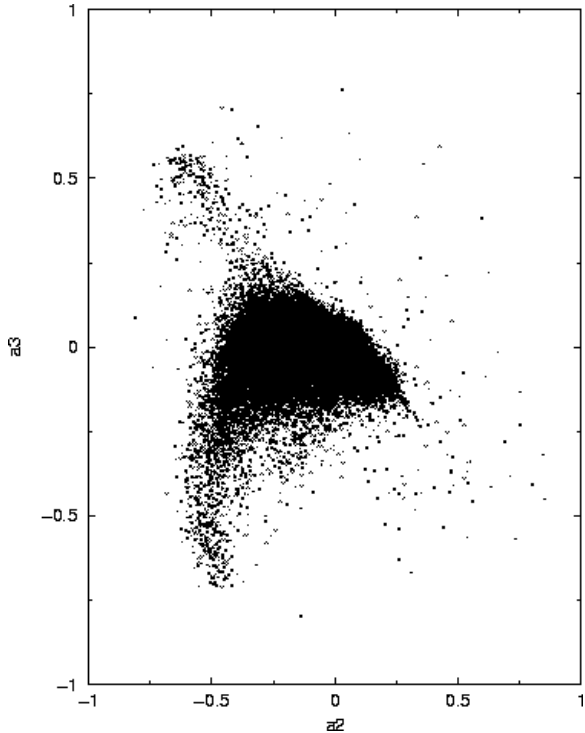}
\figcaption[a3vsa2.eps]{Eigencoefficients $a_3$ versus $a_2$ of our
sample ($\approx$170,000 galaxy spectra).  The introduction of the third
eigencoefficient further discriminates galaxies with post-starburst
activity (they are the group of galaxies with negative $a_3$ and $a_2$
values in this plot).  Also, a group of outliers are apparent (a small
group of objects with positive $a_3$ and negative $a_2$ values) which
are explained in the text.  \label{fig:a3vsa2}}
\end{figure}

\begin{figure}[h]
\includegraphics[angle=270,width=3in]{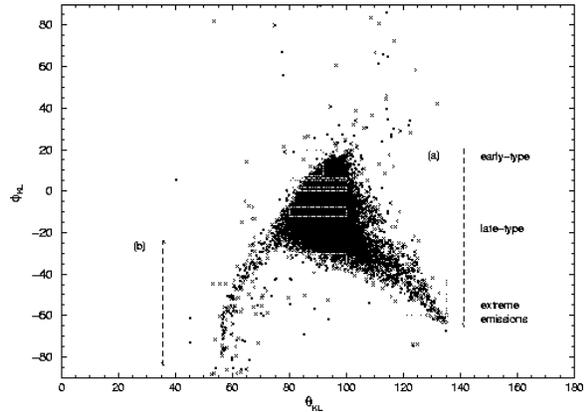}
\figcaption[phitheta_iter20.eps]{($\phi_{KL},
	\theta_{KL}$)-classification of $\approx$170,000 SDSS galaxy spectra.  (a)
	illustrates the sequence along which the galaxy spectral types
	are identified.  (b) Outliers, mostly spectra without
	significant spectral features. The angles are in degrees. Most
	outliers have large errors in their redshift estimations, while
	90\% have low signal-to-noise ratios. The boxes are areas in
	which the mean of all of the observed spectra correspond to red,
	blue and emission line galaxies. See Figure~\ref{fig:meanspec}
	for the mean spectra.  \label{fig:phitheta}}
\end{figure}

\begin{figure}[h]
\includegraphics[angle=0,width=3in]{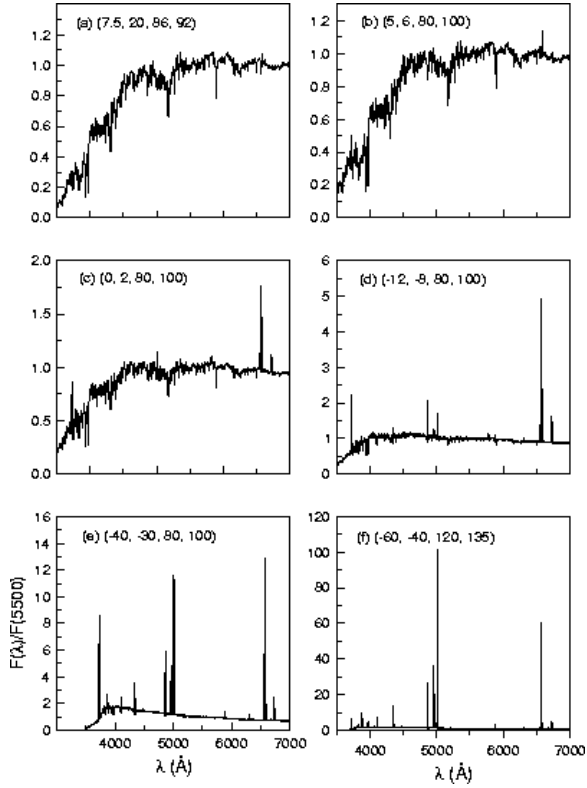}
\figcaption[meanspec_phitheta.eps]{Mean of all the observed spectra in
	different ranges of ($\phi_{KL}^{s}, \phi_{KL}^{e},
	\theta_{KL}^{s}, \theta_{KL}^{e}$), with the classification
	angles being (a) (7.5, 20, 86, 92), (b) (5, 6, 80, 100), (c) (0,
	2, 80, 100), (d) (-12, -8, 80, 100), (e) (-40, -30, 80, 100) and
	(f) (-60, -40, 120, 135) are shown.  \label{fig:meanspec}}
\end{figure}

The first three eigencoefficients of the whole sample are plotted in
Figures~\ref{fig:a2vsa1} and \ref{fig:a3vsa2} in the forms of $a_2$
versus $a_1$ and $a_3$ versus $a_2$. More than 99\% of the total galaxy
population is located on the locus in Figures~\ref{fig:a2vsa1}, in which
the second eigencoefficients have values from $\approx$ 0.25 to
-0.75. The appearance of this locus is very similar to previous works
\cite{Con95}. Red galaxies have positive, and relatively large second
eigencoefficients, while blue galaxies have smaller, or in some cases
negative values.  From Figure~\ref{fig:a3vsa2} we clearly see that by
introducing the third eigen-component, there is a group of galaxies
being separated out from the main group. These galaxies, with negative
$a_3$ and negative $a_2$ values, exhibit post-starburst activity in
their spectra.  A much smaller group (about 0.1\%) with positive $a_3$
and negative $a_2$ values is also seen. These are outliers and will be
discussed later.  The resulting $\phi_{KL}$ versus $\theta_{KL}$ is
plotted in Figure~\ref{fig:phitheta} for all galaxies in the sample
excluding those galaxies with $a_{1}<0$ (118 objects). These $a_{1}<0$
sources tend to be either artifacts within the data (M.~SubbaRao,
private communication) or spectra that are not visually confirmed as a
galaxy spectrum.  In Figure~\ref{fig:phitheta}a, the sequence from red
to blue to extreme emission line galaxies is illustrated. The boxes
drawn show the regions from which a set of spectral types are
identified.  They range from the early-type at the top of the plot to
emission line galaxies at the bottom. The spectra for these subsamples
are shown in Figure~\ref{fig:meanspec}(a-f), ranging from red to
emission line galaxies.  The spectral energy distributions shown are the
mean of all the {\it observed} spectra classified to be in the range
($\phi_{KL}^{s}$, $\phi_{KL}^{e}$, $\theta_{KL}^{s}$,
$\theta_{KL}^{e}$), where the superscripts $^s$ and $^e$ denote the
starting and ending values bounding the range. The actual values are
chosen such that the resultant mean spectra agree with the galaxy
spectra of each type in Kennicutt's atlas of nearby galaxy spectra
\cite{Kennicutt92}.  The flux levels are in good agreement with those in
the atlas, which leads us to believe that the classification is
physically sound as well as having statistical rigor.  Spectra with
similar spectral features are therefore seen to be clustered by the KL
procedure.  Due to the smoothing of spectral inhomogeneities with a
large number of galaxies, the resultant mean spectra have very high
signal-to-noise levels. This result demonstrates the power of the KL
transform for calculating mean (or composite) spectra (e.g., Eisenstein et
al.\ 2003 for the mean spectrum of the SDSS massive galaxies).

Table~\ref{tab:number} shows the number of observed galaxy spectra in
each of the regions described previously. We stress that the sum of all
galaxies listed in the table is not equal to the total number of
galaxies in the data set because the ranges chosen comprise a subset of
the full ($\phi_{KL}, \theta_{KL}$)-plane. The early late- to
intermediate late-types galaxies (with $-12^{o} < \phi_{KL} < 5^{o}$,
$80^{o} < \theta_{KL}<100^{o}$) dominate within the whole data set which
agrees with the well-known fact that late-type galaxies dominate the
field populations in terms of number counts.

Apart from the main locus in Figure~\ref{fig:phitheta} the region marked
``(b)'' identifies a group of outliers, forming approximately 0.1\% of
the full sample ($190$ sources).  These are unusual sources that arise
due to artifacts within the reduction pipelines, errors within the
spectra themselves or possibly due to new classes of astrophysical
sources. In later processing runs (idlspec2D v4.9.8, as of 13th of
August, 2002) only 68 of these sources remain in the main galaxy
sample. Approximately half of them have the ZWARNING flag set to 4,
which indicates that there are large errors in the redshift estimations.
This results in less than 0.02\% of the spectroscopic sample having
spectra that can be considered unphysical (a testament to the remarkable
accuracy and performance of the current spectroscopic reduction
pipelines).  Considering all of these sources as a whole 90\% have
signal-to-noise ratios (S/N) higher than the mean survey quality (the
$<S/N>$ is 15.9 in the data set).
 
\begin{figure}[h]
\includegraphics[angle=270,width=3in]{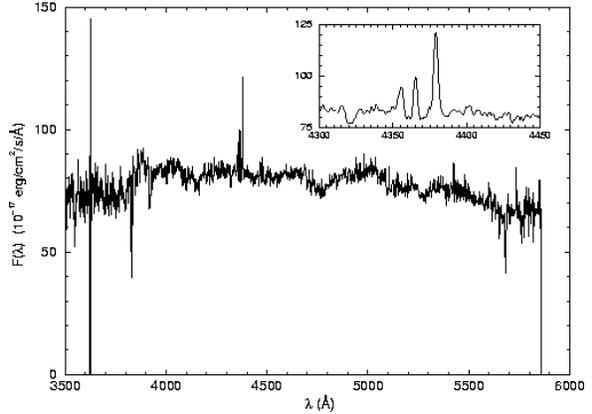}
\figcaption[weirdspec_176455.eps]{One of the outliers in the
	($\phi_{KL}, \theta_{KL}$)-plane. The redshift of this galaxy is
	incorrectly assigned by the spectroscopic pipeline.
	\label{fig:weirdspec_176455.eps}}
\end{figure}

Of the remaining 30 galaxies within this outlier class, most have
relatively high redshifts ($z\sim 0.2-0.5$) as assigned by the
pipelines. Some of these sources classed as galaxies by the pipelines do
not appear to be galaxy spectra when inspected visually. On the other
hand, for those that {\it are} galaxies as inspected by us, we found
that the pipelines have assigned incorrect redshifts to some of these
spectra. As expected, the gap-repairing procedure fails in those objects
and the resulting expansion coefficients have unphysical values.  An
example is shown in Figure~\ref{fig:weirdspec_176455.eps}, according to
the assigned redshift, this object has a redshift $0.5394$, which is
obviously incorrect from the locations of \ion{N}{2}+H$\alpha$+\ion{N}{2} lines, as
shown in the insert (this galaxy should have a redshift of $0.0236$).
The outcome is that the magnitudes of the 2nd and the 3rd
eigencoefficients obtained by a KL of all the objects in this group are
roughly the same but with different {\it signs}, meaning that no lines
that are representative of typical spectral types are found.  This
result suggests that KL technique is a powerful tool for identifying
artifacts within any spectral reduction procedure.

\begin{figure}[h]
\includegraphics[angle=0,width=3in]{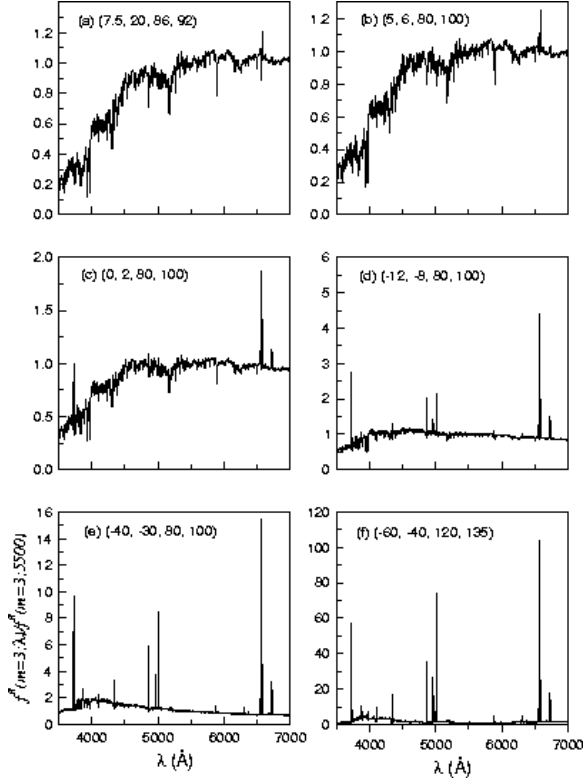}
\figcaption[reKLspec_3modes.eps]{The means of all KL-reconstructed
	spectra (a-f) in different ranges of ($\phi_{KL}^{s},
	\phi_{KL}^{e}, \theta_{KL}^{s}, \theta_{KL}^{e}$) (the bounding
	boxes are the same as those in Figure~\ref{fig:meanspec}). The
	first three eigenspectra are used in the reconstruction. The
	continua and most of the line features are in excellent
	agreement with those of the mean spectra shown in
	Figure~\ref{fig:meanspec}. \label{fig:3modes}}
\end{figure}

The above results show that the classification is successful in
allowing the galaxy types to be identified using the first three
eigencoefficients and that it may serve as a way for error checking.
How do the eigenspectra actually perform in
reconstructing the spectra?  Figure~\ref{fig:3modes}(a-f) shows, for
the same range of ($\phi_{KL}^{s}, \phi_{KL}^{e}, \theta_{KL}^{s},
\theta_{KL}^{e}$) as above, the means of all KL-reconstructed
spectra. A KL-reconstructed spectrum, using $m$-eigenspectra, is given
by
\begin{equation}
  f^R(m; \lambda)=\sum_{k=1}^{m} a_{k} e_{k}(\lambda) \ .
\end{equation}

It should be noted that the KL-reconstructed spectrum is different from
the KL-repaired spectrum we mentioned previously (in that case the
repairing is in the gap regions only).  For convenience, the {\it mean}
of all KL-reconstructed spectra in a given range is abbreviated as
``KL-reconstructed spectrum'' in the following sections unless otherwise
specified.  Comparing the mean spectra in Figure~\ref{fig:meanspec}(a-f)
with the reconstructed ones in Figure~\ref{fig:3modes}(a-f) (3 modes are
used), the continuum levels and most emission lines are in excellent
agreement with the mean spectra (except for the galaxies with extreme
emissions, which we will discuss later).  These results are consistent
with previous claims that two eigenspectra are enough to describe most
of the spectral types in galaxies \cite{Con95}.  Our present
classification scheme of using two mixing angles of the first three
eigencoefficients is also justified.

\begin{figure}[h]
\includegraphics[angle=270,width=3in]{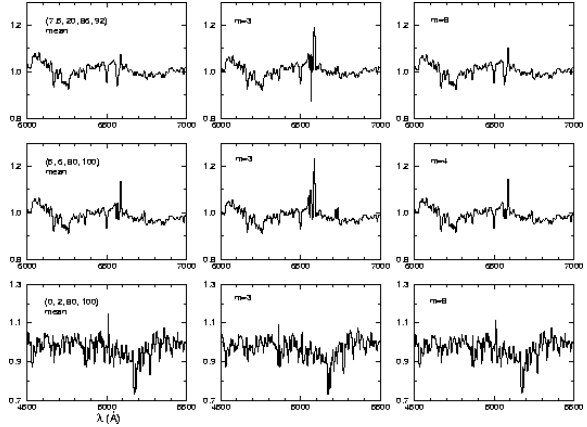}
\figcaption[recon_lines_details.eps]{Number of modes needed to
	reconstruct some of the lines in early-type to early late-type
	galaxy, with the classification angles in the ranges (7.5, 20,
	86, 92) (top panel), (5, 6, 80, 100) (middle panel) and (0, 2,
	80, 100) (bottom panel). The figures on the leftmost panels show
	the mean spectra in each type, and the consecutive figures show
	the KL-reconstructed spectra with different numbers of
	modes. All spectra are normalized at
	5500\AA.\label{fig:recon_lines_details}}
\end{figure}

\begin{figure}[h]
\includegraphics[angle=270,width=3in]{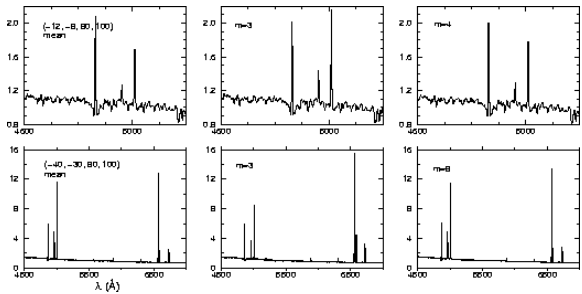}
\figcaption[recon_lines_details_2.eps]{Same as
	Figure~\ref{fig:recon_lines_details}, but for galaxies with the
	classification angles in the ranges (-12, -8, 80, 100) (top
	panel) and (-40, -30, 80, 100) (bottom
	panel). \label{fig:recon_lines_details_2}}
\end{figure}

The 3-mode KL-reconstructed spectra shown in
Figure~\ref{fig:3modes}(a-f) also suggest that to reconstruct some of
the lines and line ratios, more eigenspectra are
necessary. Figure~\ref{fig:recon_lines_details} shows in detail the
emission lines that require more than 3 modes for reconstruction. These
figures show early-type to late-type galaxies (from top to bottom). With
the first eight eigenspectra, the amplitude of the \ion{N}{2} line for galaxies
with classification angles in the range (7.5, 20, 86, 92) can be
correctly recovered.  Similarly, the first four modes are sufficient for
the \ion{N}{2} and H$\alpha$ reconstruction for galaxies with classification
angles in the range (5, 6, 80, 100). Progressing to bluer galaxies with
classification angles in the range (0, 2, 80, 100), the first eight
modes are enough for the \ion{O}{3} reconstruction.  Similarly,
Figure~\ref{fig:recon_lines_details_2} shows the cases for galaxies with
prominent emission features. We find that the first four modes are
enough to reconstruct the amplitudes and line-ratios
\ion{O}{3}[5008.240]/\ion{O}{3}[4960.295] and H$\beta$[4862.68]/\ion{O}{3}[4960.295] for
those with classification angles in the range (-12, -8, 80, 100). For
galaxies in the range (-60, -40, 120, 135), the line-ratio
\ion{N}{2}[6585.27]/\ion{N}{2}[6549.86] is correct with three eigenspectra, while the
first eight modes are enough to further retrieve the amplitudes of the
two \ion{N}{2} lines. The maximum differences between the amplitudes of the
mean and the reconstructed lines (which we define as the error of the
reconstruction of a particular line) in the above-mentioned cases are
about 10\%.

\begin{figure}[h]
\includegraphics[angle=270,width=3in]{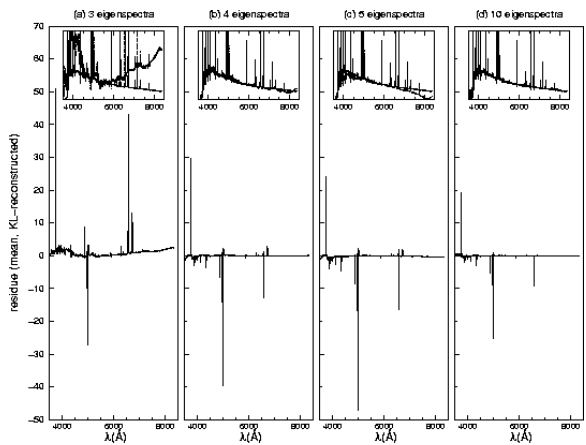}
\figcaption[SBmresidue.eps]{The residuals of the mean spectrum of
	extreme-emission galaxies (classification angles in the range
	(-60, -40, 120, 135)) with the KL-reconstructed spectrum using
	(a) 3, (b) 4, (c) 5, and (d) 10 eigenspectra. The inserts are
	the enlarged regions of the continuum levels, in each case the
	solid line is the mean spectrum and the dotted line is the mean
	of the KL-reconstructed spectra.  The spectra are normalized at
	5500\AA.\label{fig:SBmresidue}}
\end{figure}

Therefore, for all but the most extreme emission line galaxies, eight
spectral components, or modes, are sufficient to reconstruct the
spectral line ratios to an accuracy of about 10\%\ (a factor of 500 in
compression of information within the galaxy spectra). For the
reconstruction of galaxies with extreme emissions, however, the
performance is not satisfactory when using a small number of
eigenspectra. Nevertheless, ten eigenspectra are sufficient to recover
the continuum level (see the mean and KL-reconstructed spectra in the
enlarged continuum region). The residuals of the mean spectra and the
KL-reconstructed spectra are shown in Figure~\ref{fig:SBmresidue}, where
(a-d) correspond to the reconstructions with 3, 4, 5 and 10 eigenspectra
respectively. There are substantial improvements in using ten
eigenspectra, especially in nebular lines and \ion{S}{2} lines, and various
line ratios. The typical errors in the fluxes of lines remains around
$15-25$\%.

This is not a surprising result. On one hand, the result follows because
of the increasingly dominant role of lines in the higher-order modes
compared with the continuum.  On the other hand, statistics also play a
factor. The early and intermediate-type galaxies dominate the population
of galaxies while emission and extreme emission line galaxies comprise
just a few percent of the total population. Thus, galaxies with
significant emission call for more eigenspectra and higher-order modes
in their reconstructions. Besides the statistical reasons, the
inevitable variations in line-widths of emission lines make it
comparatively difficult to reconstruct them accurately using linear
combinations of eigenspectra.

Due to the fact that the spectral features in extreme emission line
galaxies are distinct from other types of galaxies, they still reveal
themselves in the plane ($\phi_{KL}, \theta_{KL}$). Thus, for the main
purpose of this work, which is obtaining a robust and objective
classification of galaxies, the less satisfactory performance of
reconstructing some emission lines in galaxies with extreme emission
lines does not have a significant effect. However, if detailed diagnosis
of lines (for example, the flux-ratio of two lines) in those galaxies
are of interest, then more modes are needed. Better yet, a separate
analysis using KL with those emission line galaxies is suggested.
 
\subsection{KL-reconstruction as Low-pass Filtering} \label{subsection:lowpassfilter}

\begin{figure}[h]
\includegraphics[angle=0,width=3in]{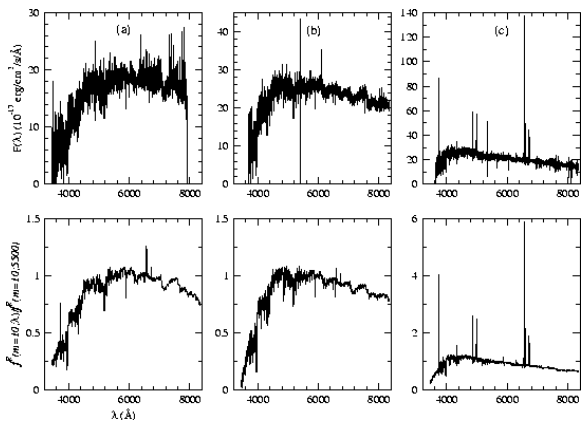}
\figcaption[inspec_reconspe_comprandom.eps]{Low-pass filtering of the
	observed spectra (a-c) by the KL-reconstruction. The figures
	on the lower panel show the KL-reconstructed spectra
	respectively for each observed spectrum (normalized at
	5500\AA).  \label{fig:inspec_reconspe_comprandom}}
\end{figure}

The inclusion of all the modes in the KL-reconstruction of a given
spectrum should, in principle, reproduce all spectral features
(including the noise). As higher-order modes contain higher frequency
signals and smaller variances of the sample, the inclusion of only a few lower-order
modes would thus suppress the noise present in the spectrum. Examples of
the comparison between the observed spectra and the KL-reconstructed
ones are shown in Figure~\ref{fig:inspec_reconspe_comprandom}(a-c). In
each case, the spectrum is reconstructed with the first 10 eigenspectra,
and normalized at 5500\AA.  From these noise-free reconstructed spectra,
it becomes a simple task to identify and classify the spectral lines.

\section{Reliability of the Galaxy Classification} \label{section:repeat}

Any classification scheme has to be repeatable in order to be useful.
If we measure the spectrum of a galaxy on different nights in different
conditions, would the classification still be the same? To answer this
question, we are fortunate in that many galaxy spectra in the SDSS data
set are taken on multiple nights \cite{Blanton02}. A total of 1,854
galaxies were found in our sample to be not unique
(i.e., they have been observed and reduced independently). A further
thirty thousand galaxies were found in the SDSS spectroscopic data 
to have been observed on multiple
nights with different observing conditions (often these individual
observations do not meet the signal-to-noise requirements of the SDSS
spectral observations). The quantitative interpretation of the
repeatability of classifications based on these plates may be difficult
due to the variation in signal-to-noise ratios.  Nevertheless all repeat
observations are selected for this part of our work.

\begin{figure}[h]
\includegraphics[angle=270,width=3in]{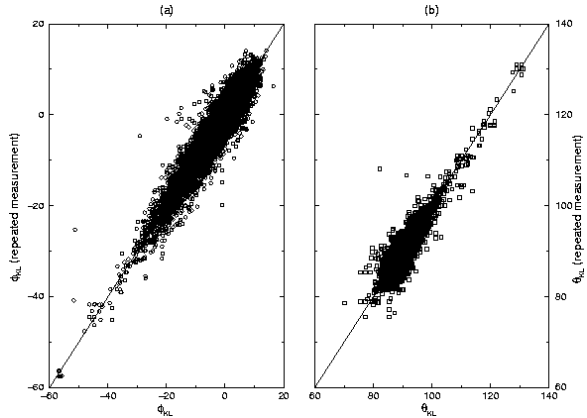}
\figcaption[matchphitheta_iter20.eps]{Reliability of the KL
	classification of galaxies. The classification parameter of
	each object is plotted against that of the repeated
	measurement, for the cases (a) $\phi_{KL}$ and (a)
	$\theta_{KL}$.  \label{fig:repeat}}
\end{figure}

Of the thirty-thousand sources, only those with flags PRIMTARGET,
OBJTYPE and CLASS equal GALAXY are selected (together with the
requirement that all sources are present in the most up to date
reductions). This selection results in thirteen-thousand objects in the
final sample. Figure~\ref{fig:repeat}a and Figure~\ref{fig:repeat}b show
a comparison between the $\phi_{KL}$ and $\theta_{KL}$ values assigned
by our classification scheme to those galaxies with the highest
signal-to-noise and the classification derived from the repeat
observations.  The solid line corresponds to the location of the
one-to-one correspondence between the two measures. The dispersions in
$\phi_{KL}$ and $\theta_{KL}$ are $2.35^{o}$ and $1.61^{o}$
respectively. This agreement is excellent, as these angle dispersions
correspond to small changes in the resulting repaired spectra. The
agreement also spans a large range in both classification angles. The
implication of this finding is that a truly reliable and repeatable
classification scheme is obtained which validates the repeatability of
the spectrophotometry of the SDSS.

\begin{figure}[h]
\includegraphics[angle=270,width=3in]{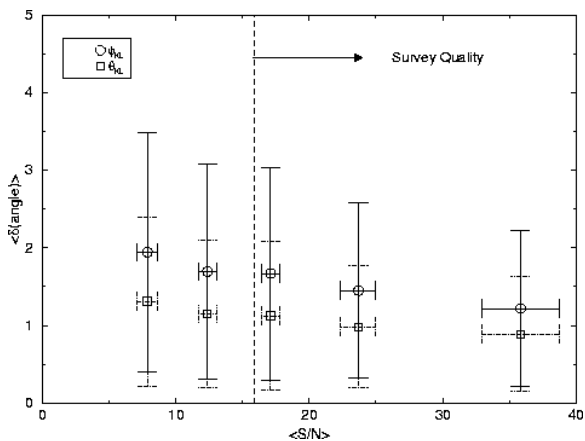}
\figcaption[SigSNphitheta.eps]{The mean discrepancy in the
	 classification angles derived from the KL analysis (circles are
	 $<|\delta\phi_{KL}|>$ and squares, $<|\delta\theta_{KL}|>$)
	 versus the mean signal-to-noise ratio of all spectra. All
	 spectra in this plot have been observed more than once.
	 \label{fig:SigSNphitheta}}
\end{figure}

In order to determine a representative signal-to-noise ratio for each
spectrum the median signal-to-noise ratio is adopted (the flag
SN\_MEDIAN in spZbest-{\it plate}-{\it mjd}.fits).  The dependence of
the rms error in the measured angles on the
signal-to-noise of the observations (where the signal-to-noise is
selected to be the lower one of any pair of observations) is shown in
Figure~\ref{fig:SigSNphitheta}.  From the current data set we observe
a weak trend with larger discrepancies in the classification for those
observations with lower signal-to-noise ratios. The mean (absolute)
discrepancies in the classification angles ($<|\delta(\phi_{KL})|>$
and $<|\delta(\theta_{KL})|>$) and mean signal-to-noise ratios are
calculated in the ranges of signal-to-noise ratio ($0.0-10.0$), ($10.0-
15.0$), ($15.0-20.0$), ($20.0-30.0$) and larger than 30.0. The dependence
is very similar in both the $\phi_{KL}$ and $\theta_{KL}$ angles.  The
error bars are set by the root-mean-square fluctuations in both
quantities. The vertical line marks the calculated mean
signal-to-noise ratio of all the galaxies defined as meeting the
survey quality (a signal-to-noise of 15.9).

For those sources meeting the survey quality signal-to-noise criteria,
the maximum errors in the two mixing angles are, approximately, three
degrees in $\phi_{KL}$ and two degrees in $\theta_{KL}$. This result
shows that the classifications based on the SDSS spectra are
repeatable and robust to the variable signal-to-noise within the
spectroscopic data. The fact that the signal-to-noise dependence is
weak suggests that the noise within the spectra are essentially
Poisson such that the projection of a noisy spectrum does not add
substantial artifacts into the expansion coefficients. Despite this
weak dependence in the distribution of expansion coefficients with
signal-to-noise we do find instances where the spectral properties of
the galaxies change between pairs of observations. For example, in one
case we find that the strength of the O~II lines change by about 20\%
between two separate observations. It is not clear whether this difference 
is due to a calibration error or due to variability in the source.

\section{A Simple Model of the Distribution of Galaxy Populations} \label{section:model}

\begin{figure}[h]
\includegraphics[angle=270,width=3in]{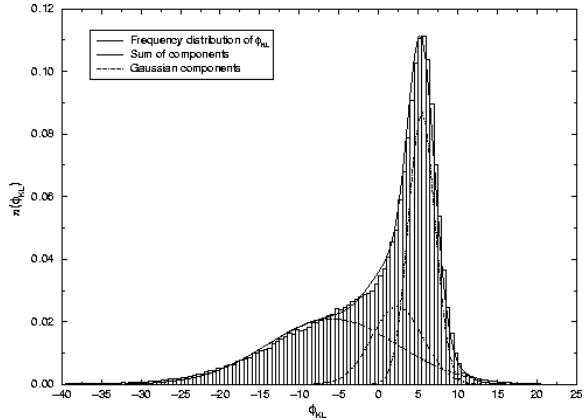}
\figcaption[out3g.eps]{Frequency distribution of $\phi_{KL}$.  The three
	modeled Gaussian populations are shown (dotted lines), with the
	peaks correspond to early late- through to intermediate
	late-types.  The solid line is the sum of the three modeled
	populations.  \label{fig:phihistogram}}
\end{figure}

From studies of the luminosity function of galaxies it has been shown
that the distribution of galaxies comprise a number of populations or
classes. It is, therefore, natural to ask how many classes are present
within the SDSS spectroscopic data and how many galaxies occupy each
class. We plot in Figure~\ref{fig:phihistogram} the frequency
distribution of $\phi_{KL}$, (for the moment we neglect $\theta_{KL}$
because the extreme emission line galaxies contribute less than a
percent to the full galaxy distribution).  The bin width is
$\phi_{KL}=0.5^{o}$ and the histogram is normalized to unity.
Visually, there appear to be two to three dominant ``classes'' or
subtypes within the $\phi_{KL}$-distribution.  To further investigate
the number of subpopulations, we adopt the Akaike Information
Criterion (AIC). AIC is widely used in model selection in a number of
different disciplines.  The details of AIC and its application in
astronomy can be found in Connolly~et~al.\ \cite{Con00}.
Basically, in AIC, a score is assigned to the model distribution,
allowing a quantitative comparison with the true distribution of the
data.  Naturally, more parameters within a model yield a better
fit to the data. To counter this the AIC penalizes the score based on
the number of parameters within a model. The AIC score is given by the
following,
\begin{equation}
{\mathrm{Score(AIC)}} \equiv  \ln{\mathcal{L}} - R
\end{equation}

\noindent
for a given model. In this definition, $\ln{\mathcal{L}}$ is the
log-likelihood function, and $R$ is the number of parameters in the
model. As a result, the higher is the score the better the model.

\begin{figure}[h]
\includegraphics[angle=0,width=3in]{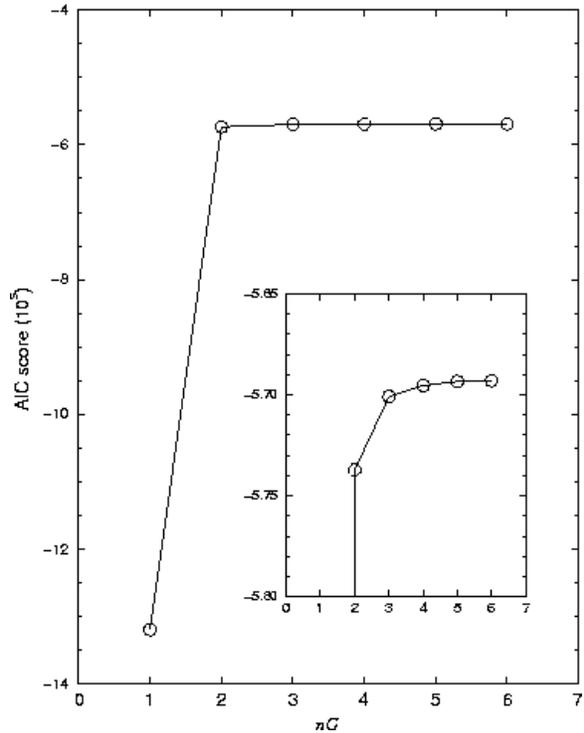}
\figcaption[AICscorePHI.eps]{AIC score ($\times 10^5$) as a function of
	the number of Gaussians in the population
	model. \label{fig:AICscore}}
\end{figure}

The first step in the analysis is to choose a functional form for
the model; a Gaussian model is adopted here.  We fit models with increasing numbers of 
Gaussian components to the $\phi_{KL}$-distribution and show the resultant 
AIC score in Figure~\ref{fig:AICscore} as a function of the number of 
Gaussians $(nG)$ in the linear mixture model. The insert shows an 
enlargement of the region around $nG=2-6$. We find that for $nG=5-6$, 
the scores start to flatten off (0.01\% difference in the AIC score), 
whereas the major improvement occurs at $nG=2$. In a statistical sense, 
$nG=5-6$ give the best score and therefore it would appear that at most six 
subgroups might contribute to the distribution of the $\phi_{KL}$
values. We do note, however, that there is no underlying physical
motivation for assuming a Gaussian mixture model and that as the
number of Gaussians in the model exceeds four the individual Gaussian
contain no direct physical meaning.  That is to say, the individual
populations for $nG>4$ actually overlap, forming redundant
descriptions. Thus, we estimate that a linear model of a mixture of 
three Gaussians is sufficient for modeling the populations of galaxies in 
our data set. The form of the model is as follows
\begin{eqnarray}
  n(\phi_{KL}) & = & G(0.087, 5.43, 2.39) + G(0.025, 2.34, 4.55) \nonumber \\
               &   & + G(0.021, -5.86, 11.58) \ ,
\end{eqnarray}
\noindent 
where $n(\phi_{KL})$ is the number density (normalized, 
$\int n(\phi_{KL}) d\phi_{KL}=1$) as a function of $\phi_{KL}$, 
and $G(C, M, S)$ is a Gaussian function of $\phi_{KL}$ 
\begin{equation}
  G(C, M, S) = C e^{{-[(\phi_{KL}-M)/S]}^2} \ .
\end{equation}

The three Gaussians are illustrated by the dotted lines in
Figure~\ref{fig:phihistogram}. Comparing the mean values (in
$\phi_{KL}$) of the Gaussian distributions with the ranges of
$\phi_{KL}$ over which the mean spectra of different galaxy types are
derived (see Figure~\ref{fig:meanspec}, also see
Table~\ref{tab:number}), these distributions correspond approximately to
early through to intermediate types. Because the first two eigenspectra,
i.e., $\phi_{KL}$, roughly describe the color of a galaxy, the different
sub-populations we obtain should relate to the color separation found in
the SDSS EDR galaxies \cite{Strateva01} in which the bimodal
$u^{*}-r^{*}$ color distribution corresponds to early- (E, S0 and Sa)
and late- (Sb, Sc and Irr) types \cite{Shimasaku01}.  Besides, optical
colors of all galaxies in the SDSS were found to be correlated very
strongly with $^{0.1}(g-r)$ color (i.e., the $g-r$ color for galaxies at
redshift $z=0.1$), which was also found to be double-peaked
\cite{Blanton03a}.

\section{Applications of KL eigenspectra} \label{section:app}

As we have shown, the eigencoefficients that describe a galaxy spectrum 
correlate strongly with its intrinsic spectral type. We will 
leave for a later paper a detailed investigation of the correlations 
inherent within the eigenbases and their relations to physical spectral 
energy distribution models such as Bruzual and Charlot (1993). In the 
following section we will just note a number of the interesting 
correlations present within the galaxy spectra and eigenspectra.

\subsection{Line Correlations within the Eigenspectra}

\begin{figure}[h]
\includegraphics[angle=270,width=3in]{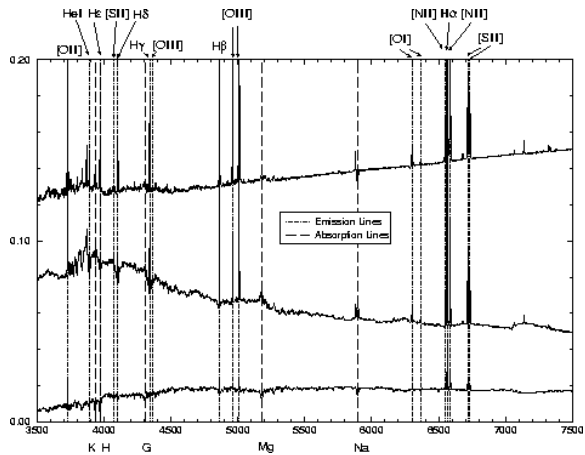}
\figcaption[eigSpec_iter20_012_LINES_names.eps]{The 1st, 2nd and 3rd
	eigenspectra overlaid on the emission and absorption lines
	identified by the SDSS spectroscopic
	pipeline.\label{fig:PClines}}
\end{figure}

Each galaxy spectrum can be constructed through a linear combination
of the eigenspectra. While the relative weights of these combinations
(i.e., the expansion coefficients) have been shown to provide a basis
for the classification of the galaxy spectra, the details of the
individual eigenspectra provide insight into the relative correlations
between the emission and absorption lines within a spectrum together
with its continuum shape. Spectral lines that are typically
anti-correlated will appear anti-correlated in the the second
eigenspectrum (e.g., one with positive emission and the other as an
absorption feature). Figure~\ref{fig:PClines} plots the first three
eigenspectra (with the first eigenspectrum the lower spectrum on the
plot) with the typical emission and absorption lines identified by the
SDSS spectroscopic pipelines \cite{EDR} overlaid.  The 2nd and 3rd
eigenspectra are flipped and offset by an arbitrary amount to improve
the clarity of this figure.

What is immediately apparent from this figure is that the majority of
the nebula lines are highly correlated.  An increase in the star
formation rate within a galaxy will result in a general increase in
the luminosity of all emission lines. While we expect this correlation
in the hydrogen lines, it does not necessarily have to be the case for
other lines such as [\ion{O}{3}]: the physical processes that give rise to
these lines are different (i.e., radiative vs collisional excitation).
The most obvious anti-correlation arises for the Na~D line at 5800~\AA. 
The first eigen-component shows the sodium absorption (commonly
associated with neutral gas at a temperature of a few thousand
degrees) to be present in the mean spectrum of galaxies. From the
second eigenspectrum, we see that as the star formation rate of the
galaxy increases (i.e., we add the second component to the mean
spectrum increasing the emission line strengths) the intensity of the
Na~D absorption line decreases. If the majority of the Na line comes
from stellar lines (averaged over the 157,000 galaxies in this sample)
then this relation is to be expected due to the increase in the
population of O stars with increasing star formation rate. A similar
relation is seen for the Mg triplet.

Considering the eigen-components individually we see that the mean
galaxy spectrum for the main galaxy sample has a significant component
in the H$\alpha$ and [\ion{N}{2}] emission lines with weaker emission from
[\ion{O}{3}] and no real evidence for Balmer emission below H$\beta$. Adding
in the second eigenspectrum has the result of increasing the overall
star formation within a galaxy (i.e., both the blue continuum increases
together with the nebular emission lines). The second eigenspectrum has
very strong Balmer absorption indicative of post starburst activity
within a galaxy. The third component is dominated by line
emission. There is very little of the stellar absorption of the Balmer
emission lines as it is seen in the second eigenspectrum. A combination,
therefore, of the first and third component will enable the
reconstruction of pure absorption or emission line spectra. Within the
third component the Ca~K and Ca~H absorption lines are strongly
anti-correlated with the emission lines. Increasing the contribution of
the third eigenspectrum has the net effect of increasing the line
emission together with decreasing the strong absorption line
features. This result is understood by the fact that absorption features
in a galaxy are mainly due to older stellar populations, and many
emission lines, especially nebular lines, are due to the ionization of
the interstellar medium within the galaxy by hot stars.

It is, therefore, clear that the correlations present within the
eigenspectra provide a reasonable description of the physical
processes that occur within typical galaxy spectra. A more detailed
description of these correlations will be the subject of a followup
paper.

\subsection{Stellar-absorption of the Hydrogen Emission Lines}

\begin{figure}[h]
\includegraphics[angle=270,width=3in]{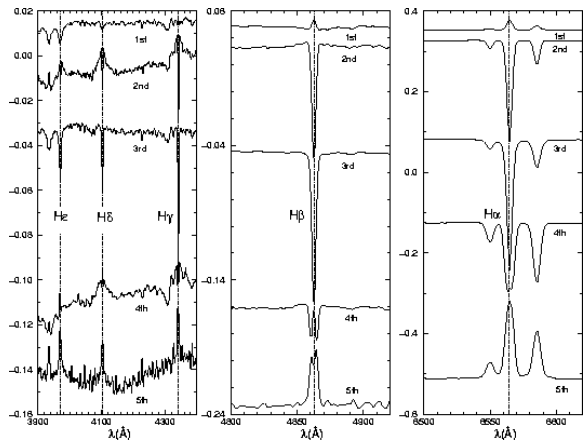}
\figcaption[eigSpec_selfabs_all.eps]{Stellar absorptions of hydrogen
	emission lines present in the eigenspectra.  The eigenspectra
	are arbitrary shifted  for clarity.
	\label{fig:eigSpec_selfabs_all}}
\end{figure}

Perhaps the most striking feature within these spectra is that the
second eigenspectrum shows the hydrogen emission lines in H$\epsilon$,
H$\delta$, H$\gamma$, and H$\beta$ exhibiting stellar absorption. The
clarity of this effect comes from the high resolution of the SDSS
spectroscopic data (relative to other large spectroscopic samples such
as the 2dF) together with the accurate control we have on the
spectrophotometric calibration of the individual spectra.
Figure~\ref{fig:eigSpec_selfabs_all} shows an enlarged region of
interest for the first four eigenspectra. Comparatively, H$\beta$ is
the weakest in terms of this effect, while H$\alpha$ shows no apparent
effect. The absorption features are also observed in higher-order modes,
but they are not shown here since the first few modes dominate. We find
that the majority of the signal for the stellar absorption comes from
the second eigen-component. There is a smaller contribution from the
fourth component but the contribution from this component describes the
variation in the widths of the hydrogen lines rather than their
amplitudes.

As we have shown in Figure~\ref{fig:3modes} using just
three modes we can recover galaxy spectra with and without strong
stellar absorption. The consequences of this are two-fold. The fourth
eigen-component appears not to contribute significantly to the stellar
absorption signal, as noted above. Secondly, the fact that just two modes
can recover the shape of the stellar absorption suggests that
the mechanism that describes the magnitude of this process is, on
average, relatively simple (as would be expected given the correlation
between the spectral properties and stellar composition of the
galaxies). This would imply that modeling the stellar absorption and
correcting for its effect on the emission line properties of galaxies
should be a straightforward process in a statistical sense, even in
the presence of low signal-to-noise data.

\section{The Effect of a Fixed Aperture} \label{section:aperturebias}

The SDSS uses a fixed aperture of 3\arcsec \, diameter for its
spectroscopic observations. This can, in principal, lead to biases in
the current spectral classification scheme if, for example, a fiber
samples only the central bulge of a nearby intermediate or late-type
galaxy (resulting in the assignment of an early-type spectral class).
As this effect depends on the apparent size of a galaxy when compared to
the fiber diameter it has the potential to induce redshift and
luminosity dependent biases in any analysis using the KL classifications
\cite{Kochanek00}. Studies of the effect of aperture bias on observed
parameters (e.g., star-formation rate) can be found in, e.g., Baldry et
al.\ (2002), P\'erez-Gonz\'alez et~al.\ (2003) and Brinchmann et~al.\ (2003).
The questions we address here are: $(i)$ is there an aperture bias using
the KL approach $(ii)$ how can we quantify this bias and $(iii)$ can we
correct for the aperture effects to obtain bias-free galaxy types?


We estimate the effect of aperture bias by calculating, for a given
galaxy, the difference in the classification (in this case the
$\phi_{KL}$ angle) derived from the total galaxy flux compared to that
derived from the central 3 arcseconds. The dependence of this
classification error on the redshift and the physical size of the galaxy
serves to quantify the bias in our sample. We assume that the apparent
diameter of each galaxy can be approximated as twice the Petrosian
half-light radius (petro50) in the r-band. The physical sizes of the galaxies are then
calculated by assuming $\Omega_m =0.3$, $\Omega_\Lambda = 0.7$ and $H_0 =
71$. The aperture magnitudes of all galaxies are initially k-corrected
to redshift $z=0.1$ using the code by Blanton et~al.\ (2003b) version
1.16 prior to estimating the spectral types. Type assignment for the
total flux and fiber flux is performed using the photometric redshift
code of Connolly et~al.\ (1999b). The input spectral templates are
constructed as linear-combinations of the first 3 eigenspectra from this
work, with the resolution in both $\phi_{KL}$ and $\theta_{KL}$ set to
$2^{o}$.  In the following discussion we will express the distance
dependence of the relation as function of $z/z_{max}$, where $z_{max}$
is the highest redshift at which a galaxy of a given absolute magnitude
would pass the sample selection criteria. This provides a pseudo volume
independent analysis.

\begin{figure}[h]
\includegraphics[angle=0,width=3in]{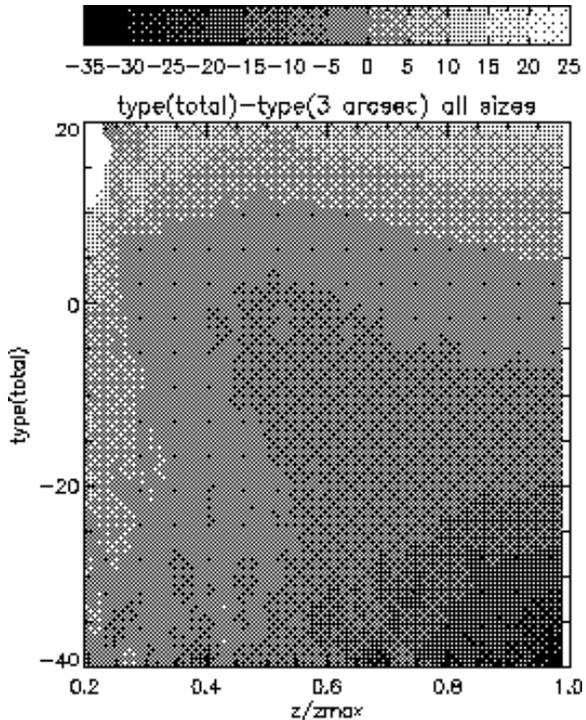}
\figcaption[dphiallgalcontour.eps]{The greyscale contour of the
        difference in the classification ($D\phi_{KL}$) 
	for the total flux and for the
        inner 3\arcsec \, region  as a function of
        $z/z_{max}$. The ordinate is the type assigned from the total flux. 
        \label{fig:dphiallgalcontour}}
\end{figure}

Figure~\ref{fig:dphiallgalcontour} shows the difference in the
classifications of galaxies, $\phi_{KL}$(total)$-\phi_{KL}$(3\arcsec)
($\equiv D\phi_{KL}$), as a function of $z/z_{max}$ and galaxy
type. The bin sizes of smoothing are 0.02 in $z/z_{max}$ and
 $2^{o}$ in $D\phi_{KL}$. Galaxies of sizes from $0-100$ kpc are included, whereas galaxies
of $\phi_{KL}$(total)$ < -40^{o}$ are excluded for there are less then 1\% of them. 
Lighter components in the greyscale image correspond to the fraction of galaxies that would
be classified as an earlier type (i.e., redder) if the total flux was
used rather than the 3 arcsec flux. Darker components correspond to
galaxies that are of later type (bluer) when using the total flux.  
The percentages of galaxies residing within
these contours are listed in Table~\ref{tab:aperturebiasNUM}. 
From our repeatability test, the mean signal-to-noise limited classification is
$<|\delta(\phi_{KL})|>=2.35$. With the assumption that the typical
signal-to-noise limit in the $\phi_{KL}$ angle estimation for the whole
galaxy is the same as that for the inner 3\arcsec, the derived
signal-to-noise limit in $D\phi_{KL}$ is $2\times <|\delta(\phi_{KL})|>
\approx 5$. There are about half of the galaxies ($\approx$ 40\%) in our sample in which
the type-differences are within the estimated signal-to-noise limit.

Aperture effects on the spectral classification clearly exist. For blue
galaxies (i.e., $\phi_{KL} \approx 0^{o} - -40^{o}$), $D\phi_{KL}$ 
increases for nearby galaxies. This is to be expected as the flux from
the inner 3 arcsec is more likely to be dominated by the presence of a
bulge component.  Similarly, for galaxies classified as red based on
their total flux (i.e., $\phi_{KL} \approx 0^{o} - 20^{o}$), errors in
the classification angles $D\phi_{KL}$ increase rapidly with decreasing
distance (i.e., $z/z_{max} < 0.25$). This implies that the cores of red
galaxies are redder than the color estimated from the total flux.

\begin{figure}[h]
\includegraphics[angle=0,width=3in]{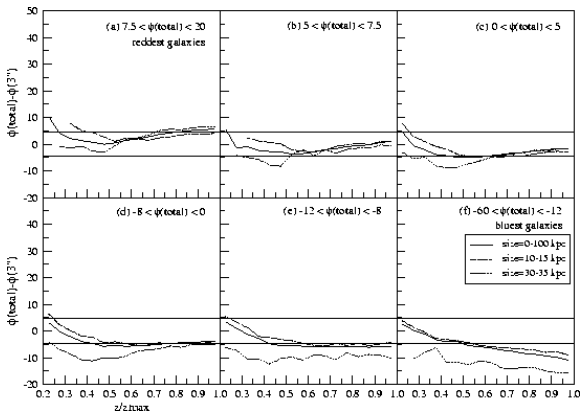}
\figcaption[dphivszoverzmax6types.eps]{The difference in the
        classification for the whole galaxy and the inner 3\arcsec \,
        region as a function of $z/z_{max}$ from the reddest (a) to the
        bluest (f) galaxies. In each sub-figure the galaxies are of
        sizes $0-100$~kpc (solid line), $10-15$~kpc (dotted line) and $30-35$~kpc
        (dashed line).  \label{fig:dphivszoverzmax6types}}
\end{figure}

The dependence of the aperture bias on the physical size of a galaxy is
illustrated in Figure~\ref{fig:dphivszoverzmax6types}. We divide the
above sample into 6 ranges of galaxy type, from the reddest in
Figure~(a) to the bluest in Figure~(f).  The differences in the
classification angles $D\phi_{KL}$ are plotted in each figure as a
function of $z/z_{max}$ for physical sizes ranging from $0-100$ kpc
(black line), $10-15$ kpc (dotted line) and $30-35$ kpc (dashed
line). The two horizontal lines mark the uncertainty on the
classification due to the survey signal-to-noise limits. For the red
galaxies in Figure~(a) to (c) the bias is constant or decreases with
effective distance and is, essentially, negligible when compared to the
uncertainties on the classification.  For distances $z/z_{max} < 0.25$,
the bias is above the signal-to-noise limit so that the type deduced
from the total flux is redder than that from the central 3\arcsec.

As we would expect Figures~(b) and (c) show a dependence on galaxy size
for the classifications with larger galaxies exhibiting a redder
classification when only considering the 3 arcsec flux. This size and
redshift dependence extends to the blue galaxies (Figure~(d) through (f)).
For these galaxies, however, a more pronounced dependency is shown on
the classification bias with galaxy size. Overall there is a general
aperture bias for all physical sizes of galaxy ($0-100$ kpc) that
approaches the intrinsic error on the classification as the effective
redshift $z/z_{max}$ approaches unity.  The exception to this arises
when we consider galaxies with prominent emission lines (Figure~(f))
which, counter-intuitively, exhibit {\it larger} bias the more distant
they are.  One of the possible reasons for this is our use of three
eigenspectra in constructing the spectral templates, whereas 10 modes
and more are typically required to accurately reconstruct these observed
spectra.

\begin{figure}[h]
\includegraphics[angle=0,width=3in]{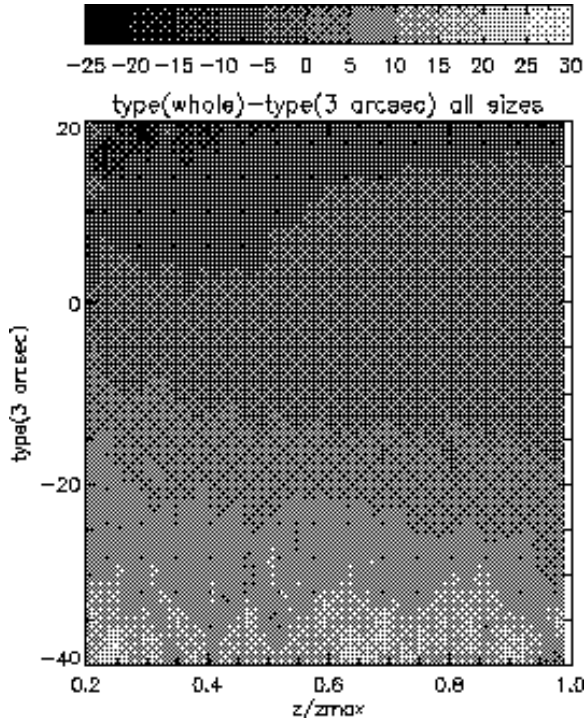}
\figcaption[dphiallgalcontour3arcsec.eps]{Same as
        Figure~\ref{fig:dphiallgalcontour} except that the ordinate is
        the type for the inner 3\arcsec \, of each galaxy.  This serves
        as the look-up table with which the aperture bias can be corrected.
        \label{fig:dphiallgalcontour3arcsec}}
\end{figure}

While we observe an aperture bias in the SDSS sample it is relatively
small when compared to the intrinsic classification errors and is
essentially negligible for most galaxies. Moreover its dependence on
size and redshift is relatively mild and straightforward to correct.
For those galaxies with non-negligible bias a simple
correction can be made using the lookup table shown in
Figure~\ref{fig:dphiallgalcontour3arcsec}, which is identical to
Figure~\ref{fig:dphiallgalcontour} with the exception that the
ordinate axes is the spectral type inside 3\arcsec,
$\phi_{KL}(3\arcsec)$. Given the $\phi_{KL}(3\arcsec)$ and
$z/z_{max}$ the correction to our spectral classification can be
determined directly from Figure~\ref{fig:dphiallgalcontour3arcsec} with 
the size-averaged bias.


\section{Conclusions and Outlook}\label{section:conclusion}

From the application of Karhunen-Lo\`eve transform, an objective
classification of $\approx$170,000 galaxy spectra in the SDSS is performed.  With
a quantitative convergence criteria defined, gappy galaxy spectra can be
repaired and KL eigenspectra and eigencoefficients derived. For most of
the galaxy types, three eigenspectra are sufficient for describing the
continua and emission lines to a high degree of accuracy with a maximum
error in line-reconstructions of approximately 10\%. Typically ten modes
are needed in the reconstruction of galaxies with extreme emission lines
with errors of $15-25$\% in the line fluxes.  We find that a two-parameter
$(\phi_{KL}, \theta_{KL})$-classification scheme can discriminate
between spectra corresponding to all spectral types used in the current
classification scheme (including galaxies with extreme emission
lines). This classification is robust to repeat observations (at a level
of a few degrees in the classification angles) due to the accurate
spectrophotometric calibration of the SDSS data set. We find a weak
dependence in the classification on the signal-to-noise of the
spectra. This effect is, however, smaller than the typical dispersion
between repeat observations and is negligible at signal-to-noise levels
at which the SDSS spectra are defined as being of survey quality.

We find that there exists a minimum number of randomly selected spectra
that are necessary to statistically represent the information within the
full sample (i.e., to be representative of the true distribution of
galaxies).  For a set of ten eigenspectra (i.e., ten eigenspectra enable
the reproduction of both quiescent and active galaxies) the number of
spectra required is around $3000$ to $4000$. This is due to the need to
sample a minimum number of randomly selected galaxies in order to
include galaxies with extreme emission line properties in our data set
(as they comprise only 0.1\% of the full galaxy sample).

We find that the bias on the spectral classification due to the fixed
aperture spectroscopy is, on average, small and is negligible for all
galaxies except for the reddest galaxies that are very close by
($z/z_{max} < 0.3$) and for those galaxies that are large physically ($>
30$ kpc) with prominent emission lines. A look-up table is constructed
for the correction of this bias.

There are several future directions related to this work. With the
present continuous classification scheme, which simplifies the
distribution of galaxies into a handful of parameters, studies of the
statistics of the physical properties of galaxies become more tractable.
The clustering and spectral properties of these classifications will be
addressed in a future paper. The generality of these techniques are
applicable to any set of spectra and has been recently applied to the
SDSS QSO catalog (Yip et~al.\ 2003, 2004).
 
\acknowledgements

We thank Mariangela~Bernardi, Ravi~Sheth and Michael~Blanton for
comments and discussions. CWY is partially supported by
Zaccheus Daniel Fellowship. AJC and CWY acknowledge partial support from an NSF
CAREER award AST99 84924, a NASA LTSA grant NAG5 8546 and a NSF ITR
award 0121671.

Funding for the creation and distribution of the SDSS Archive has been
provided by the Alfred P. Sloan Foundation, the Participating
Institutions, the National Aeronautics and Space Administration, the
National Science Foundation, the U.S. Department of Energy, the
Japanese Monbukagakusho, and the Max Planck Society. The SDSS Web site
is http://www.sdss.org/.

The SDSS is managed by the Astrophysical Research Consortium (ARC) for
the Participating Institutions. The Participating Institutions are The
University of Chicago, Fermilab, the Institute for Advanced Study, the
Japan Participation Group, The Johns Hopkins University, Los Alamos
National Laboratory, the Max-Planck-Institute for Astronomy (MPIA),
the Max-Planck-Institute for Astrophysics (MPA), New Mexico State
University, University of Pittsburgh, Princeton University, the United
States Naval Observatory, and the University of Washington.

\clearpage

\begin{table}[h]
\begin{center}
\begin{tabular}{|c||c|c|c|c|}
\hline
 ($\phi_{KL}^{s}, \phi_{KL}^{e}, \theta_{KL}^{s}, \theta_{KL}^{e}$) & 
Galaxy Type & Number [Number ratio relative to Sa] \\
\hline
\hline
    (7.5, 20, 86, 92) & E/S0   & 6599  [0.34]
 \\
\hline
      (5, 6, 80, 100) & Sa     & 19543 [1.00]
 \\
\hline
      (0, 2, 80, 100) & Sb     & 13872 [0.71]
 \\
\hline
   (-12, -8, 80, 100) & Sbc/Sc & 11979 [0.61]
 \\
\hline  
  (-40, -30, 80, 100) & Sm/Im  & 140   [0.0072]
 \\
\hline
 (-60, -40, 120, 135) & SBm    & 135   [0.0069]
 \\
\hline
\hline
\end{tabular}
\end{center}
\caption{The number of galaxies in the range ($\phi_{KL}^{s},
\phi_{KL}^{e}, \theta_{KL}^{s}, \theta_{KL}^{e}$). These data are a
subset of the full sample.  The galaxy types listed are the possible
morphological types, estimated by comparing the spectral features of the
mean spectrum constructed in each range with spectra in
\cite{Kennicutt92} and therefore they are for reference only.
\label{tab:number}}
\end{table} 

\clearpage

\begin{table}
\begin{center}
\begin{tabular}{|r|c|}
\hline
 $m$ (number of modes) & weight (normalized by total weight) \\ 
\hline
\hline
   1 & 0.9594 \\
   2 & 0.9784 \\
   3 & 0.9815 \\
   5 & 0.9837 \\ 
   8 & 0.9849 \\
  10 & 0.9852 \\
  20 & 0.9855 \\
  50 & 0.9860 \\
 100 & 0.9867 \\
 500 & 0.9908 \\
1000 & 0.9940 \\
\hline
\end{tabular}
\end{center}
\caption{The relative weights of the eigenspectra. The first 3
eigenspectra comprise about 98\% of the total sample variance.
\label{tab:weight}}
\end{table}

\begin{table}
\begin{center}
\begin{tabular}{|c|c|}
\hline
 $D\phi_{KL}$-range & fraction of galaxies \\ 
 (degree) & (0.1\% accuracy) \\ 
\hline
\hline
$ -20 < D\phi_{KL} <= -10  $ & 18.5 \% \\ 
$ -10 < D\phi_{KL} <= -5   $ & 16.4 \% \\ 
$  -5 < D\phi_{KL} <= -2.5 $ & 9.2  \% \\ 
$-2.5 < D\phi_{KL} <= 0    $ & 17.7 \% \\ 
$   0 < D\phi_{KL} <= 2.5  $ & 7.3  \% \\ 
$ 2.5 < D\phi_{KL} <= 5    $ & 6.1  \% \\ 
$   5 < D\phi_{KL} <= 10   $ & 12.1 \% \\ 
$  10 < D\phi_{KL} <= 20   $ & 7.2  \% \\ 
\hline
\end{tabular}
\end{center}
\caption{The number of galaxy spectra in our sample with the specified values of aperture bias $D\phi_{KL}$.
\label{tab:aperturebiasNUM}}
\end{table}


\clearpage
 
\begin{thebibliography}{}

\bibitem[Bailer-Jones et~al.\ 1998]{Bailer-Jones98} Bailer-Jones~C.~A.~L., 
Irwin~M. \& von~Hippel~T. \ 1998, MNRAS, {\bf 298}, 361.

\bibitem[]{} Baldry~I.~K.,Glazebrook~K.~B.,
 Carlton~M. et~al. \ 2002, ApJ, {\bf 536}, 68.

\bibitem[Blanton et~al.\ 2002]{Blanton02} Blanton~M.~R., Lupton~R.~H., 
Maley~F.~M. et~al. \ 2002, submitted to AJ.

\bibitem[Blanton et~al.\ 2003a]{Blanton03a} Blanton~M.~R., Hogg~D.~W., 
Bahcall~N.~A. et~al.\ 2003a, ApJ, {\bf 594}, 186.

\bibitem[Blanton et~al.\ 2003b]{Blanton03b} Blanton~M.~R., Brinkmann~J.,
 Csabai~I. et~al.\ 2003b, AJ, {\bf 125}, 2348.

\bibitem[Boroson \& Green 1992]{Boroson92} Boroson~T.~A. \& Green~R.~F. \ 1992, 
ApJ Supp. Series, {\bf 80}, 109.

\bibitem[]{} Brinchmann~J., Charlot~S., White~S.~D.~M. et~al. \ 2003, submitted to MNRAS.

\bibitem[Bromley et~al.\ 1998]{Bromley98} Bromley~B.~C., Press~W.~H., 
Lin~H. \& Kirshner~R.\ 1998, ApJ, {\bf 505}, 25.

\bibitem[Castander et~al.\ 2001]{Castander01} Castander~F.~J., Nichol~R.~C.,
Merrelli~A. \ 2001, AJ, {\bf 121}, 2331.

\bibitem[]{coll01} Colless~M.~M., Dalton~G., Maddox~S. et~al. \ 2001, 
MNRAS, {\bf 328}, 1039.

\bibitem[2000]{Con00} Connolly~A.~J., Genovese~C., Moore~A.~W. et~al. \ 2000, submitted to AJ.

\bibitem[Connolly et~al.\ 1995]{Con95} Connolly~A.~J., 
Szalay~A.~S., Bershady~M.~A., Kinney~A.~L. \& Calzetti~D.\ 1995, 
ApJ, {\bf 110}, 1071.

\bibitem[Connolly \& Szalay 1999a]{Con99a} Connolly~A.~J. 
\& Szalay~A.~S.\ 1999a, ApJ, {\bf 117}, 2052.

\bibitem[Connolly \& Szalay 1999b]{Con99b}
Connolly~A.~J., Budav\'ari~T., Szalay~A.~S., Csabai~I. \& Brunner~R.~J. \
1999b, in ASP Conf. Ser. 191, {\it Photometric Redshifts and High
Redshift Galaxies}, ed. Weymann~R.~J., Storrie-Lombardi~L.~J.,
Sawicki~M., \& Brunner~R.~J. (San Francisco: ASP), 13.

\bibitem[Efstathiou \& Fall 1984]{EfstathiouFall84} Efstathiou~G. \& 
Fall~M.~S.\ 1984, MNRAS, {\bf 206}, 453.

\bibitem[]{}Eisenstein~D.~J., Annis~J., Gunn~J.~E. et~al. \ 2001, AJ, {\bf 122}, 
2267.

\bibitem[Eisenstein et~al.\ 2003]{Eisenstein03} Eisenstein~D.~J. Hogg~D.~W.,
 Fukugita~M. et~al. \ 2003, ApJ, {\bf 585}, 694.

\bibitem[Everson \& Sirovich 1994]{Everson94} Everson~R. \& Sirovich~L.\
1994, J.Opt.Soc.Am.A, {\bf 12}, No.8, 1657.

\bibitem[Folkes et~al.\ 1996]{Folkes96} Folkes~S.~R., Lahav~O. \& Maddox~S.~J.\ 
1996, MNRAS, {\bf 283}, 651.

\bibitem[Folkes et~al.\ 1999]{Folkes99} Folkes~S.~R., Ronen~S., Price~I. et~al.\ 1999, 
MNRAS, {\bf 308}, 459.

\bibitem[Francis et~al.\ 1992]{Francis92} Francis~P.~J., Hewett~P.~C., Foltz~C.~B. 
\& Chaffee~F.~H\ 1992, ApJ, {\bf 398}, 476.

\bibitem[]{} Fukugita~M., Ichikawa~T., Gunn~J.~E. et~al. \ 1996, AJ, {\bf 111}, 1748.

\bibitem[Galaz \& de Lapparent 1998]{Galaz98} Galaz~G. \& de Lapparent~V.\ 1998, A\&A, 
{\bf 332}, 459. 

\bibitem[]{} Gunn~J.~E., Carr~M., Rockosi~C. et~al. \ 1998, AJ, {\bf 116}, 3040.

\bibitem[]{} Hogg~D.~W., Finkbeiner~D.~P., Schlegel~D.~J. \& Gunn~J.~E. \ 2001, 
AJ, {\bf 122}, 2129.

\bibitem[Hubble 1926]{Hubble26} Hubble~E.\ 1926, ApJ, {\bf 64}, 321.

\bibitem[Kennicutt 1992]{Kennicutt92} Kennicutt~R.~C.,~Jr.\ 1992, ApJ 
Supp. Series, {\bf 79}, 255.

\bibitem[Kochanek et~al.\ 2000]{Kochanek00} Kochanek~C.~S., Pahre~M.~A. \& Falco~E.~E.
astro-ph/0011458.

\bibitem[]{} Lupton~R., Gunn~J.~E., Ivezi\'c~Z., Knapp~G.~R. \&
Kent~S.\ 2001, in ASP Conf. Ser. 238, {\it Astronomical Data Analysis
Software and Systems X}, ed. Harnden~F.~R.~Jr., Primini~F.~A., \&
Payne~H.~E. (San Francisco: Astr. Spc. Pac.), p. 269.

\bibitem[Merzbacher 1970]{Merzbacher70} Merzbacher~E.\ 1970, 
{\it Quantum Mechanics}, second edition, John Wiley\&Sons.

\bibitem[Murtagh \& Heck 1987]{Murtagh87} Murtagh~F. \& Heck~A.\ 1987,
{\it Multivariate data analysis}, Astrophysics and Space Science Library No 131, 
Dordrecht: Reidel.

\bibitem[]{} P\'erez-Gonz\'alez~P.~G., Zamorano~J., Gallego~J., Arag\'on-Salamanca~A. \&
Gil~de~Paz~A. \ 2003, ApJ, {\bf 591}, 827.

\bibitem[]{} Pier~J.~R., Munn~J.~A., Hindsley~R.~B. et~al. \ 2002, AJ, {\bf 125}, 1559.

\bibitem[]{} Richards~G.~T., Fan~X., Newberg~H.J. et~al. \ 2002, AJ, {\bf 123}, 2945.

\bibitem[Ronen et~al.\ 1998]{Ronen98} Ronen~S., 
Arag\'on-Salamanca~A. \& Lahav~O. \ 1999, MNRAS, {\bf 303}, 284.

\bibitem[Shimasaku et~al.\ 2001]{Shimasaku01} Shimasaku~K., Fukugita~M.,
 Doi~M. et~al. \ 2001, AJ, {\bf 122}, 1238.

\bibitem[Singh et~al.\ 1998]{Singh98} Singh~H.~P., Gulati~R.~K. 
\& Gupta~R. \ 1998, MNRAS, {\bf 295}, 312.

\bibitem[]{} Smith~J.~A., Tucker~D.~L., Kent~S. et~al. \ 2002, AJ, {\bf 123}, 
2121.

\bibitem[Sodre \& Cuevas 1997]{Sodre97} Sodr\'e~L.~Jr \& Cuevas~H. \ 1997, 
MNRAS, {\bf 287}, 137.

\bibitem[Stoughton et~al.\ 2002]{EDR} Stoughton~C.,  Lupton~R.~H.,
 Bernardi~M. et~al. \ 2002, ApJ, {\bf 123}, 485.

\bibitem[Strauss et~al.\ 2002]{maingalaxy} Strauss~M.~A., Weinberg~D.~H., 
Lupton~R.~H. et~al. \ 2002, AJ, {\bf 124}, 1810.

\bibitem[Strateva et~al.\ 2001]{Strateva01} Strateva~I., Ivezic~Z.,
 Knapp~G.~R. et~al. \ 2001, ApJ, {\bf 122}, 1861. 

\bibitem[Yip et~al.\ 2003]{Yip03} Yip.~C.~W., Connolly~A.~J.,
Vanden Berk~D.~E. et~al. \ 2003, in ASP Conf. Ser. 311,
{\it AGN Physics with the Sloan Digital Sky Survey},  ed. Hall~P.~B. \& Richards~G.~T.. 

\bibitem[Yip et~al.\ 2004]{Yip04} Yip.~C.~W., Connolly~A.~J.,
Vanden Berk~D.~E. et~al. \ 2004, in prep (SDSS Publication No. 331). 

\bibitem[York et~al.\ 2000]{SDSS} York~D.~G., Adelman~J.,
 Anderson~J.~E.~Jr. et~al. \ 2000,
ApJ, {\bf 120}, 1579. 

\end{thebibliography}
\end{document}